\documentclass[longauth]{aa}  

\usepackage{graphicx}
\usepackage{txfonts}
\usepackage[]{hyperref}
\usepackage{color}
\hypersetup{colorlinks=true,allcolors=[rgb]{0,0,0.8}}

\usepackage{lineno}

\usepackage{multirow}
\usepackage{adjustbox} 

\usepackage{natbib}
\bibpunct{(}{)}{;}{a}{}{,} 

\makeatletter
\renewcommand*\aa@pageof{, page \thepage{} of \pageref*{LastPage}}
\makeatother

\newcommand{\Msun}{\ensuremath{M_\sun}}

\newcommand{\TESS}{\emph{TESS}}

\newcommand{\kms}{km\,s$^{-1}$}

\newcommand{\logl}{\ensuremath{\log(L/L_\sun)}}

\newcommand{\mj}{\ensuremath{\,M_{\rm J}}}

\newcommand{\msun}{\ensuremath{M_\sun}}

\newcommand{\rsun}{\ensuremath{R_\sun}}

\newcommand{\teff}{$T_{\rm eff}$}

\newcommand{\vsini}{$v$sin$i$}

\begin{document}

   \title{A planetary-mass candidate imaged in the\\Young Suns Exoplanet Survey}


 \author{Pengyu Liu 
          \inst{1,2,3}
          \and
          Matthew A. Kenworthy\inst{1}
          \and
          Beth A. Biller\inst{2,3}
          \and
          Alex Wallace\inst{4}
          \and
          Tomas Stolker\inst{1}
          \and
          Sebastiaan Haffert\inst{1,5}
          \and
          Christian Ginski\inst{6}
          \and
          Eric E. Mamajek\inst{7,8}
          \and
          Alfred Castro-Ginard\inst{1}
          \and
          Tiffany Meshkat\inst{9}
          \and
          Mark J. Pecaut\inst{10}
          \and
          Maddalena Reggiani\inst{11}
          \and
          Jared R. Males \inst{5}
          \and
          Laird M. Close \inst{5}
          \and
          Olivier Guyon \inst{5,12,13,14}
          \and
          Isabella Doty \inst{15}
          \and
          Kyle Van Gorkom \inst{5}
          \and
          Alex Hedglen \inst{16}
          \and
          Maggie Kautz \inst{12}
          \and
          Jay Kueny \inst{12}
          \and
          Joshua Liberman \inst{12}
          \and 
          Jialin Li \inst{5}
          \and
          Joseph D. Long \inst{17}
          \and
          Jennifer Lumbres \inst{5}
          \and
          Eden McEwen \inst{12}
          \and
          Logan Pearce \inst{5}
          \and
          Roswell R. Roberts IV \inst{5}
          \and
          Lauren Schatz \inst{18}
          \and
          Katie Twitchell \inst{12}        
          }

   \institute{Leiden Observatory, Leiden University, PO Box 9513, 2300 RA Leiden, The Netherlands\\
              \email{pengyu.liu@ed.ac.uk}
        \and
            SUPA, Institute for Astronomy, University of Edinburgh, Royal Observatory, Blackford Hill, Edinburgh EH9 3HJ, UK
        \and
            Centre for Exoplanet Science, University of Edinburgh, Edinburgh, UK
        \and    
            School of Physics and Astronomy, Monash University, Victoria 3800, Australia
        \and
            Steward Observatory, University of Arizona, 933 North Cherry Avenue, Tucson, AZ 85719, USA
        \and
            School of Natural Sciences, Centre for Astronomy, University of Galway, Galway H91 CF50, Ireland
        \and
            Jet Propulsion Laboratory, California Institute of Technology, 4800 Oak Grove Drive, M/S 321-100, Pasadena CA 91109, USA
        \and 
            Department of Physics and Astronomy, University of Rochester, Rochester NY 14627, USA
        \and
            IPAC, California Institute of Technology, M/C 100-22, 1200 East California Boulevard, Pasadena, CA 91125, USA
        \and 
            Rockhurst University, Department of Physics, 1100 Rockhurst Road, Kansas City MO 64110, USA
        \and
            Institute of Astronomy, KU Leuven, Celestijnenlaan 200D, B-3001 Leuven, Belgium
        \and
            Wyant College of Optical Science, University of Arizona, 1630 E University Blvd, Tucson, AZ 85719, USA
        \and
            National Astronomical Observatory of Japan, Subaru Telescope, National Institutes of Natural Sciences, Hilo, HI 96720, USA
        \and
            Astrobiology Center, National Institutes of Natural Sciences, 2-21-1 Osawa, Mitaka, Tokyo, JAPAN
        \and
            Department of Mechanical Engineering, Hopeman Building, University of Rochester, Rochester, New York
        \and
            Northrop Grumman Corporation, 600 South Hicks Road, Rolling Meadows, Illinois
        \and
            Center for Computational Astrophysics, Flatiron Institute, 162 5th Avenue, New York, New York
        \and
            Air Force Research Laboratory, Directed Energy Directorate, Space Electro-Optics Division, Starfire Optical Range, Kirtland Air Force Base, NM, 87117,USA
             }

   \date{Received 12 February 2024; accepted 15 May 2025}

 
  \abstract
   {Directly imaged exoplanets in wide orbits challenge current gas giant formation theories. 
   They need to form quickly and acquire enough material before the disk dissipates, which cannot be accommodated by in-situ formation by core accretion.}
   {We search for wide separation ($>100$\,au) planetary-mass companions with the Young Suns Exoplanet Survey (YSES).
   Here, we present a planetary-mass candidate companion discovered in the survey.}
   {We conducted follow-up observations of the candidate system after the first epoch observations and obtained six epochs of observations for this system between 2018 and 2024, and integral field spectroscopy of the stellar component.}
   {We report the detection of a candidate companion with $H=22.04\pm0.13$ mag at a projected separation of 730 $\pm$ 10\,au away from the primary star.
   High angular resolution imaging observations of the central star show it is a visual binary.
   Acceleration data, orbital fitting, spectral energy distribution fitting and radial velocity differences all suggest that there is at least one more unresolved low-mass stellar companion in this system.
   The planetary-mass candidate shows a significant proper motion comparable to that of the primary star.
   We estimate an age of 19--28\,Myr for the primary star. We cannot confirm the companionship of the candidate due to the unknown barycentre of the stars.}
  {Long-term imaging and radial velocity monitoring of the central stars, along with spectroscopy of the candidate companion, are key to resolving the nature of this system. If confirmed, the candidate companion would have a mass of 3--5\,\mj{} estimated with the ATMO evolutionary model. It would be another cold low-mass planet imaged similar to 51~Eri~b and AF~Lep~b.
  Its extremely wide separation from the host star would challenge the formation theory of gas giant exoplanets.}


   \maketitle
%

\section{Introduction}
The number of directly imaged exoplanets has grown steadily over the past decade owing to imaging surveys with ground-based facilities.
About 40\% of the imaged exoplanets are at separations larger than 100\,au from their host stars.
For instance, AB~Pic~b is a 13.5 $\pm$ 0.5 \mj{} companion of a K0 star at a projected separation of 275\,au \citep{Chauvin2005};
COCONUTS-2b is a 6.4 $\pm$ 2 \mj{} cool companion of a field M dwarf at a projected separation of 6471\,au \citep{ZhangZ2021b};
2MASS~J1155-7919~b is a 10 \mj{} comoving companion of a M3 dwarf at a projected separation of 582\,au \citep{DicksonVandervelde2020};
HD~203030b is an 8--15 \mj{} companion at a projected separation of 487\,au orbiting around a G8 star \citep{Miles-Paez2017};
2M~2236+4751~b is an 11--14 \mj{} companion at a projected separation of 230\,au orbiting around a K7 star \citep{Bowler2017};
GU~Psc~b is a 9--13 \mj{} companion at a projected separation of 2000\,au orbiting around a young M3 star \citep{Naud2014}.
The discovery of these exoplanets challenges the formation theories of gas giant exoplanets.

It is not clear how these planets are being formed so far from the places we understand gas giant exoplanets to form.
There are two widely accepted scenarios for the formation of gas giant exoplanets: core accretion \citep[e.g.][]{Pollack1996} and gravitational instability \citep[e.g.][]{Boss1997}.
Core accretion is based on the formation of a rocky embryo close to the ice line within a circumstellar disk.
Once the embryo reaches a critical mass, runaway accretion of gas continues until the zone around the planet is cleared.
However, core accretion is problematic for wide-separation planets.
It takes more than $10^7$\,yr for core formation by runaway planetesimal accretion beyond 5\,au \citep{Goldreich2004}, which is longer than the gas dissipation time of the disk.
Therefore, planet migration is required for wide-separation planets if they are formed by core accretion \citep{Dodson-Robinson2009}.
Instead of core accretion by planetesimals, core growth by pebble accretion can form planets beyond 50\,au much faster than traditional core accretion \citep{Lambrechts2012}.
However, it still requires ejection or scattering processes to explain the observed extremely wide-separation planets.
Gravitational instability is driven by the initial fragmentation of the protostellar gas cloud.
This formation mechanism has a faster timescale than that of core accretion and forms planets typically at 20 to 200\,au from the star, supporting in-situ planet formation.
But for planets at even wider separations, it also requires planet migration.
In addition, gravitational capture of free-floating planets is another possible formation mechanism for wide-separation planetary systems \citep{Perets2012}.

There are several direct imaging surveys to search for exoplanets, such as the Spectro-Polarimetric High-contrast Exoplanet REsearch \citep[SPHERE,][]{Beuzit2019} infrared survey for exoplanets \citep[SHINE,][]{Chauvin2017} and the Gemini Planet Imager (GPI) Exoplanet Survey \citep[GPIES,][]{Macintosh2015}.
Among these surveys, the Young Suns Exoplanet Survey has a relatively high planet detection yield \citep[YSES,][]{Bohn2019, Bohn2020a, Bohn2020c, Bohn2021a}.
It is a direct-imaging snapshot survey of exoplanets targeting a homogeneous sample of 70 solar-mass pre-main-sequence stars in the Lower Centaurus Crux (LCC) subgroup of the Scorpius–Centaurus association \citep[Sco-Cen,][]{deZeeuw1999, Pecaut2016} with SPHERE.
It has detected two wide-separated exoplanets orbiting around a K3 star so far: YSES~1bc, a 14 $\pm$ 3 \mj{} companion at a projected separation of 160\,au and a 6 $\pm$ 1 \mj{} companion at a projected separation of 320\,au \citep{Bohn2020a, Bohn2020c}.

In this work, we present the detection of a planetary-mass candidate companion of 2MASS J10065573-6352086 (hereafter: 2M1006) as part of YSES. The primary star is referred to as star A and the fainter star is referred to as star B.
The observations of the system and data reduction are detailed in Sect.~\ref{sec:obs}.
In Sect.~\ref{sec:star}, we present the analysis of the central stars, a stellar system consisting of a young Sun analogue and an M dwarf. We find tensions between all the data and speculate that there is at least one more stellar component in this system.
The photometric and astrometric analysis of the planetary-mass candidate are presented in Sect.~\ref{sec:astrophot}.
We discuss the probability of the candidate being a free-floating object and comoving planet in Sect.~\ref{sec:discussion} and summarise this work in Sect.~\ref{sec:summary}.

\section{Observations and data reduction}
\label{sec:obs}
We obtained six epochs of observations of 2M1006 between 2018 and 2024 with SPHERE and MagAO-X, including imaging and integral field unit (IFU) spectra.
We summarise the imaging observation setup and conditions in Table~\ref{tab: obs_summary}.

\begin{table*}
 \centering
 \caption{Imaging observation setup and observing conditions.}
  \begin{tabular}{llllllll}
  \hline\hline
 Date & MJD & Mode & Filter & DIT(s)$\times$NDIT$\times$N\tablefootmark{a} & $\textup{T}_\textup{exposure}$ (min) & Field rotation ($^{\circ}$) & Seeing ('')\\
 \hline 
 2018-11-15 & 58437.3110 & CI & $H$ & 32$\times$1$\times$4 & 2.13  &  0.46  &  0.54 \\
 2018-11-15 & 58437.3153 & CI & $K_S$ & 32$\times$1$\times$4  & 2.13  &  0.46  &  0.54  \\
 2021-12-02  & 59550.2894 &  CI & $H$ &  32$\times$1$\times$8   & 4.26  &  1.11  &  0.70  \\
 2021-12-26  & 59574.2406 &  DBI & $K12$ & 96$\times$1$\times$16  & 25.60  &  0  & 0.71 \\
 2021-12-26 & 59574.2406 & DBI & $J23$ & 96$\times$1$\times$4  & 6.40 &  0 &  0.64 \\
 2023-01-28  & 59972.1385 &  CI & $K_S$ & 32$\times$2$\times$16  & 17.07  &  9.62 &  1.16   \\
 2023-03-03  & 60006.0451 &  CI & $H$ & 32$\times$4$\times$8  & 17.07  &  10.21  &  0.69 \\
 2024-03-23 & 60392.1920 & MagAO-X/Sci & $z'$ &0.25$\times$240 & 1.00 & 0 & 0.59 \\
\hline
\hline\end{tabular}
\tablefoot{
\tablefoottext{a}{DIT: detector integration time; NDIT: number of DIT; N: number of exposures.}
}
\label{tab: obs_summary}
\end{table*}

\subsection{SPHERE Observations and data reduction}
The SPHERE instrument is an extreme adaptive optics system and coronagraphic facility mounted on the Unit Telescope 3 of the ESO Very Large Telescope (VLT) \citep{Beuzit2019}.
2M1006 was observed with SPHERE the infrared dual-band imager and spectrograph \citep[IRDIS,][]{Dohlen2008} in five epochs on 2018-11-15, 2021-12-02, 2021-12-26, 2023-01-28, and 2023-03-03.
Observations in the first, second, fourth, and fifth epoch were obtained in the classical imaging mode (CI) with the broad-band $H$ and $K_S$ filters, while observations in the third epoch were obtained in the dual-band imaging mode \citep[DBI,][]{Vigan2010} using the $K$12 and $J$23 narrow band filter combinations. 
Only observations in the third epoch used the field-tracking mode, while observations in other epochs used the pupil-tracking mode.
All the observations used a coronagraph (N\_ALC\_YJH\_S) to mask the central star.

The central star was resolved to be composed of two stars in the first epoch.
To measure the precise location of the primary star and fainter star, we performed point spread function (PSF) fitting to the binary in the flux image where the central star is not masked.
We built a PSF template from other stars in YSES in the broad $H$ and $K_S$ band, and then fitted two PSFs simultaneously to the binary using the \textsc{nautilus} package \citep{Lange2023}, which implements importance sampling and efficient space exploration using neural networks.
The two stars were well resolved in the first epoch, marginally resolved in the second epoch, and moderately resolved in the fifth epoch.
To better fit the location, we fixed the flux ratio obtained from the first epoch for the PSF fitting in the second and fifth epochs.
For the third epoch in $J$2 and $J$3 we scaled the PSF template in the $H$ band by the corresponding wavelength and performed the fitting. 
There are multiple non-coronagraphic images (flux images) of the central stars in each epoch. We fitted the binary in each individual frame and took the mean and standard deviation as the fitting results and uncertainty of each epoch. The initial uncertainties of the position and the flux ratio are on the order of $10^{-2}$ pixels and $10^{-3}$. 
We performed astrometric calibration following the latest version of SPHERE User Manual\footnote{\url{https://www.eso.org/sci/facilities/paranal/instruments/sphere/doc.html}} and \cite{Maire2021}.
To correct the systematic error between the PSF template and the observed images, we empirically included an additional 1 mas in RA and Dec in the final position uncertainty and the root mean square (rms) of the residuals in the final ratio uncertainty. This uncertainty agrees with the typical binary measurements in \citep{Bonavita2022} which do not use PSF fitting.
Table~\ref{tab:error_table} lists the error budget for the binary fitting. The final uncertainty is calculated by the propagation of error.

\begin{table*}[ht]
\begin{center}
\caption{Astrometry error budget of the central binary and the candidate.}
\begin{tabular}{lllllll}
\hline \hline
Object & Fitting (mas) & Pixel scale (mas) & True north (deg) & Pupil offset (deg) & Distortion (mas) & Systematics (mas) \\
\hline
Binary & $\sim$0.5 & $\sim$0.1 & 0.08 & 0.11 & $\sim$0.02 & $\sim$1 \\
Candidate & $\sim$3 & $\sim$7 & 0.08 & 0.11 & $\sim$2 & $\sim$1 \\
\hline
\end{tabular}
\label{tab:error_table}
\end{center}
\end{table*}

In science images where the central star is blocked by the coronagraph, the binarity of the central source challenges image alignment between the flux images and science images.
For coronagraphic observations of SPHERE/IRDIS, there are four waffle spots in the centred image, which are duplications of the PSF of the central star dispersed with wavelength \citep{Langlois2014, Beuzit2019}. %
They are used to determine the precise location of the central star behind the coronagraph by the intersection of the two pairs of spots.
The positions of the four spots are fitted by a 2D-Gaussian model, which is effective when the central star is a single star.
However, when the central star is a blend of two sources, it is problematic to use a single 2D-Gaussian fitting, because the spots are dispersed images of the two sources.
To measure the precise location of the brighter star in the coronagraphic images, we developed two methods to tackle this problem. 
The first one is to fit two 2D-Gaussian models to the spot simultaneously.
The separation and flux ratio of the binary obtained from the PSF fitting of the flux image were taken as initial guesses for the two 2D-Gaussian fittings.
After measuring the positions of the sources in the four waffle spots, we measured the positions of the brighter star by the intersection of the two pairs of the spots corresponding to the primary star.
The second method is to perform cross-correlation between a simulated waffle image of the blended source and the observed image.
We built an intensity profile of waffle spots of a single star and convolved it with the flux image.
Then we cross-correlated the simulated waffle image with the observed waffle image.
The peak position of the cross-correlation function is the offset between the two images. 

The telescope was supposed to be pointed at the primary star.
There are two sets of flux images taken before and in the middle of the science observations in the fifth epoch.
We centred on the primary star and derotated the flux images to the orientation where North is up and East is left.
The fainter star is at the same position in the derotated flux images, verifying that the telescope was pointed at the primary star.
After locating the primary star in the coronagraphic images, we derotated the coronagraphic images back to the North centred on the primary star.
Then we stacked the derotated images by averaging. 
We applied angular differential imaging \citep[ADI,][]{Marois2006} and principal component analysis \citep[PCA,][]{Soummer2012,Amara2012} to the derotated images.
No objects were detected at separations smaller than 1\farcs6.
Because the candidate is at a wide separation, we took the stacked image as the final reduced image.

\begin{figure*}
\centering
  \includegraphics[width=18.5cm]{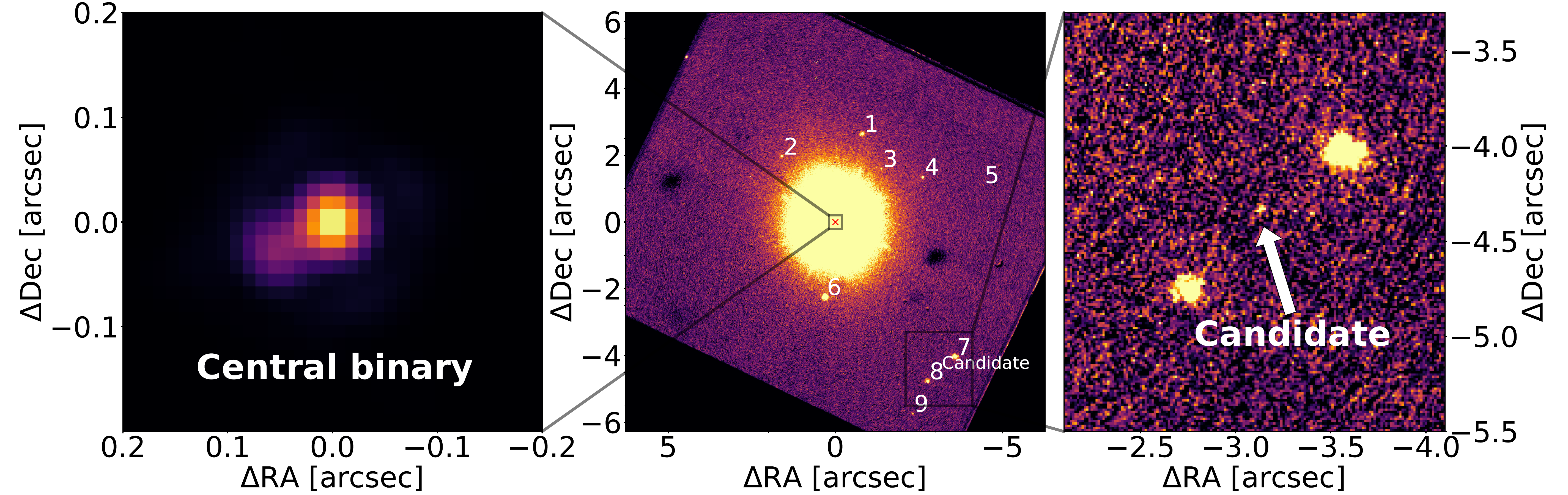} 
\caption{Full image of this system of the first epoch in the $H$ band.
Left panel: central stars.
The fainter component is in the southeast direction of the primary star on 2018-11-15.
Middle panel: stacked coronagraphic image centred on the primary star.
There are ten sources in the field of view in total.
The planetary-mass candidate is the faint source between the two bright background stars in the southwest.
Right panel: zoomed-in image of the candidate.
All images are derotated to the orientation where North is up and East is left.
} 
\label{fig:wholeime}
\end{figure*}

\subsection{MagAO-X Observations and data reduction}
The six-epoch observations were taken with the MagAO-X instrument \citep{males2022magao}.
MagAO-X is an extreme adaptive optics instrument for the Magellan Clay telescope at Las Campanas Observatory.
MagAO-X has been specifically designed for direct imaging at visible wavelengths \citep{close2018optical}.
The system was observed on 2024-03-23 in the $z^\prime$ filter with the EMCCD camera camsci1 on MagAO-X.
The observations were made in direct imaging mode to perform follow-up astrometric measurements of the tight binary system.
One minute of further observations were taken in pupil tracking mode.
No significant field rotation occurred during this time, and therefore the images were stacked before any processing was done.
After stacking, basic detector calibration was performed by subtracting a dark frame.
The data were then centred on the primary star.
It was not possible to create a PSF model directly from the data due to the small separation of the two objects.
YSES~2 was observed directly before the observations of this system.
Those observations were used to create a reference PSF because of the targets' similarity in elevation, seeing conditions and brightness during the MagAO-X observations.
The YSES~2's PSF was centred to create the reference PSF.
Both components of 2M1006 were fitted with the reference PSF.
A true north correction of 1\fdg57 $\pm$ 0\fdg2 \citep{Long2025} was applied.

\subsection{MagAO-X/VIS-X Observations and data reduction}
Spectroscopic data were taken with the MagAO-X system.
MagAO-X has an integral-field unit, the Visible Integral-field Spectrograph eXtreme \citep[VIS-X; ][]{haffert2022visible}.
VIS-X is a microlens-based IFU and it has been designed for high-spectral resolution ($R=13,500=\Delta v=20\mathrm{km s}^{-1}$) observations of the H$\alpha$ emission line.
The spectral range is limited to about 5~nm around the H$\alpha$ line due to the microlens design.
2M1006 was observed with VIS-X to measure the radial velocity difference between the two stars.
The VIS-X observations were taken on 2024-04-02.
The total on-target time is 40 minutes, consisting of 8 exposures of 5 minutes each.
The seeing condition varied between 0\farcs6 and 0\farcs75.

The VIS-X data reduction follows a typical IFU data reduction pipeline (Haffert et al. in prep).
First, the detector calibration steps are applied by subtracting a dark frame.
Internal source calibrations were used to identify the spectral trace model for each spaxel.
This model can then be used to extract the 1D spectra for all the spaxels.
The shifts in the wavelength solution between spaxels were calibrated using standard star observations, which was the star Sirius in this case.
The IFU calibration is completed after these steps.
The second part of the data reduction follows a similar approach to the imaging data.
The data cubes were derotated and a true north correction was applied.
The spectra extraction is described in Sect.~\ref{sec:radial_velocity}.

\section{Stellar properties}
\label{sec:star}
2M1006 was identified to be a candidate member of the LCC subregion of Sco-Cen \citep{Preibisch2008, Pecaut2016}.
Gaia DR3 reports a parallax of $\varpi = 7.3241\pm0.0689$ mas \citep{Brown2021}, which translates to an estimated distance of $d = 136.15^{+1.66}_{-1.58}$\,pc \citep{BailerJones2021}.
The first spectroscopic study of the star came from \citet{Torres2006} who reported it as a K0V(e) star with strong Li absorption (EW(Li I 6707) = 350\,m\AA), EW(H$\alpha$) = 0.0\,\AA, fast projected rotation ($v \sin i$ = $77.0\pm7.7$ \kms) and a radial velocity of 14.0 \kms.
\citet{Kiraga2012} reported time-series photometry from the ASAS-South survey, and reported variability with a period $P$ = 0.7271 d.
TESS observed it in Sector 10, 11, 36, 37, 38, 63 and 64.
It shows a distinct rotation modulation of 0.73\,d in all these sectors. \citet{Fetherolf2023} reports detection of a rotation period of $P_{rot}$ = $0.725893\pm0.00557$ day in the \TESS\, Sector 10 data), agreeing with the variability period reported by the ASAS-South survey.

2M1006 was resolved as two stars in SPHERE observations in Fig.~\ref{fig:wholeime}. Given their small separation, they most likely form a bound binary pair. After considering multiple lines of evidence (detailed in the following sections), there is also significant evidence for at least one more stellar companion in the system.

\subsection{Acceleration}
A companion can affect the proper motion of a star through gravitational reflex motion, and therefore, a proper motion anomaly (PMA) which is the difference between the proper motion measured in the long term and the proper motion measured in the short term is an indicative sign of a perturbing companion \citep{Kervella2019}.
To search for any PMA of 2M1006, we checked the Hipparcos \citep{Perryman1997}, Tycho-2 \citep{Hog2000}, and UCAC5 catalogues \citep{Zacharias2017}.
This star is not in the Hipparcos catalogue but is in the Tycho-2 and UCAC5. The difference in proper motion between Tycho-2 and Gaia is within the large uncertainty of Tycho-2.  
The proper motion provided in the UCAC5 catalogue is a long-term proper motion, calculated as the difference in position between the mean UCAC epoch and the mean Gaia DR1 epoch, divided by the difference in time between those mean epochs. The long-term proper motion can be assumed to follow the barycentre of the system and to be relatively independent of perturbations from a companion \citep{Kervella2019}. The proper motion provided in the Gaia catalogue is a short-term proper motion, over the epochs covered in the Gaia DR chosen. The difference between the short-term proper motion and the long-term proper motion is the proper motion anomaly (PMA) and is an indicator of the degree to which an unseen companion causes the measured short-term proper motion to deviate from the barycentric proper motion. We thus construct the Gaia-UCAC5 PMA as:
\begin{equation}
PMA = \mu_{Gaia} - \mu_{UCAC5-Gaia}
\end{equation}
where $\mu_{Gaia}$ is the reported short-term Gaia proper motion (short-term) and $\mu_{UCAC5-Gaia}$ is the reported long-term proper motion between the UCAC and Gaia DR1 epoch.

The star's $\mu_\alpha \cos \delta$ is -21.5 $\pm$ 1.2 mas yr$^{-1}$ and $\mu_\delta$ is 8.1 $\pm$ 1.2 mas yr$^{-1}$ between UCAC5 and Gaia DR1. Its $\mu_\alpha \cos (\delta)$ is -17.606 $\pm$ 0.086 mas yr$^{-1}$ and $\mu_\delta$ is 4.892 $\pm$ 0.079 mas yr$^{-1}$ in Gaia DR3. The proper motion difference is larger than 1$\sigma$, suggesting that a fainter companion is perturbing its on-sky motion.
Moreover, although Gaia DR3 identifies 2M1006 as a single star, its Renormalised Unit Weight Error (RUWE) is 6.012, indicating a poor astrometric fit for a single-star model, which for single stars have values close to unity.
Additionally, two flare events in Tess Sector 36 were observed in 2021 with an interval of $\sim$14\,d, which could be due to the fainter star, a likely M dwarf. There was another strong flare event at the end of the Sector 36 observations, but it was only captured at the very beginning of the flare, which increased the flux by 10\%.

\subsection{Orbital fitting}
\label{sec:starorbit}
Table~\ref{tab: binary-astrometry} lists the relative astrometry and flux ratio of the fainter star to the primary star measured in our observations.
The flux ratio of the fainter star to the primary star varies between 0.18 and 0.34 from $z^\prime$ to $K_S$, indicating a later spectral type for the fainter star.
The images of the stars and residuals after fitted star subtraction can be found in Appendix~\ref{app:starims}.
The fainter star moves from the southeast to the northeast relative to the primary star from 2018 to 2024.

\begin{table*}
\centering
\caption{Astrometric and photometric measurements of the fainter star relative to the primary star.}
\begin{tabular}{llllll}
\hline \hline
Epoch (jyr) & Filter & Sep (mas) & PA (deg) & Flux ratio & $\Delta$Mag (mag)\\
\hline
2018.8715 & $H$ & 57.22$\pm$1.07 & 116.33$\pm$1.06 & 0.34$\pm$0.02 & 1.17 $\pm$ 0.07 \\ 
2018.8715 & $K_S$ & 57.52$\pm$1.21 & 117.54$\pm$1.21 & 0.33$\pm$0.02 & 1.19 $\pm$ 0.06 \\
2021.9186 & $H$ & 31.58$\pm$1.25 & 41.58$\pm$2.37 & - & - \\
2021.9842 & $J$2 & 31.65$\pm$1.38 & 35.84$\pm$2.52 & 0.33$\pm$0.04 & 1.21 $\pm$ 0.14 \\
2021.9842 & $J$3 & 32.23$\pm$1.42 & 35.80$\pm$2.54 & 0.32$\pm$0.04 & 1.24 $\pm$ 0.15\\
2023.1664 & $H$ & 39.81$\pm$1.06 & 5.40$\pm$1.55 & - & - \\
2024.2237 & $z^\prime$ & 54.20$\pm$0.62 & 349.51$\pm$0.65 & 0.18$\pm$0.02 & 1.88 $\pm$ 0.12\\
\hline
\label{tab: binary-astrometry}
\end{tabular}
\centering
\end{table*}

\begin{figure*}
\centering
  \includegraphics[width=0.8\textwidth]{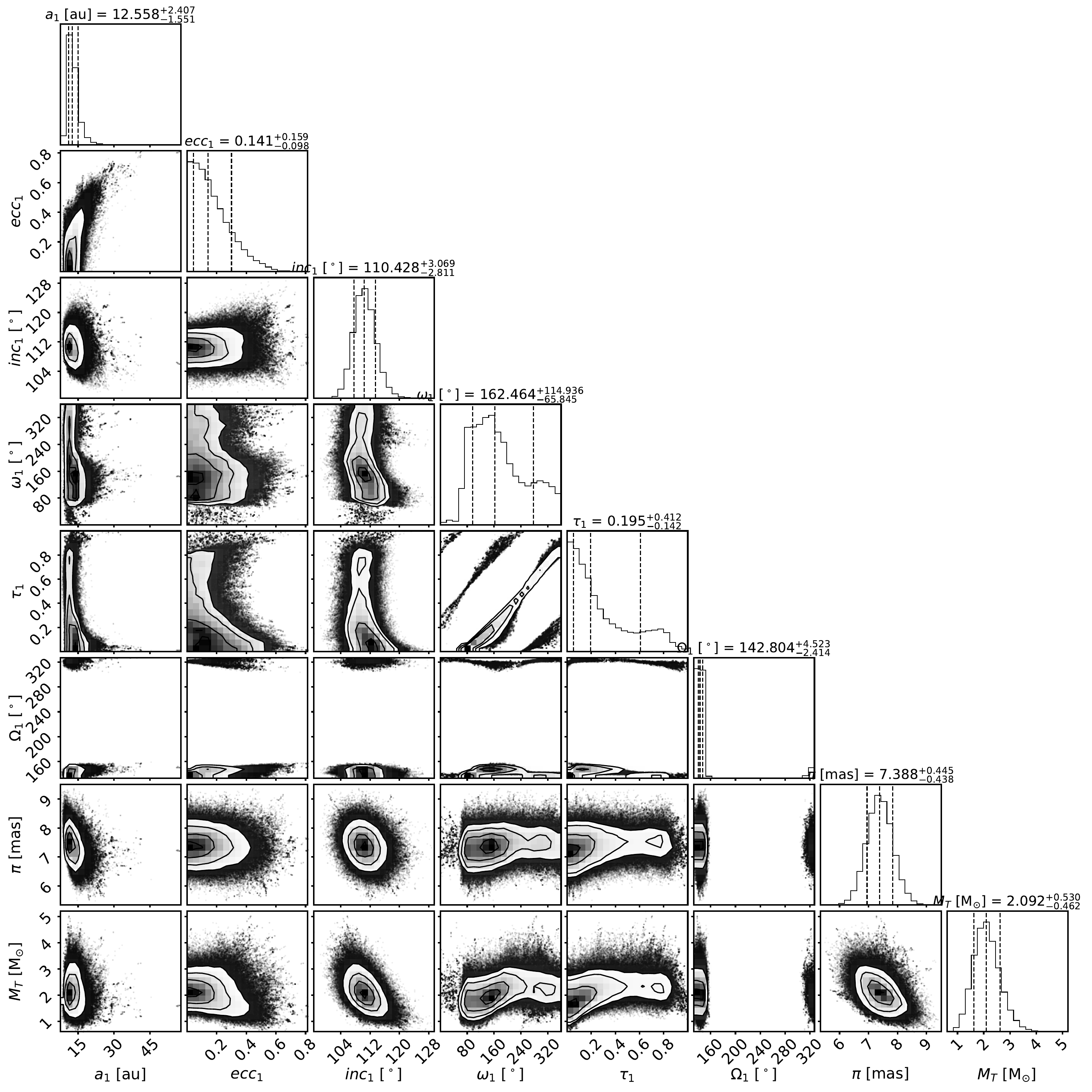}  
\caption{Posterior distributions of the orbital parameters by fitting the measured relative positions with \textsc{orbitize}.
} 
\label{fig:binary_orbitize_corner}
\end{figure*}

\begin{figure*}
\centering
  \includegraphics[width=0.8\textwidth]{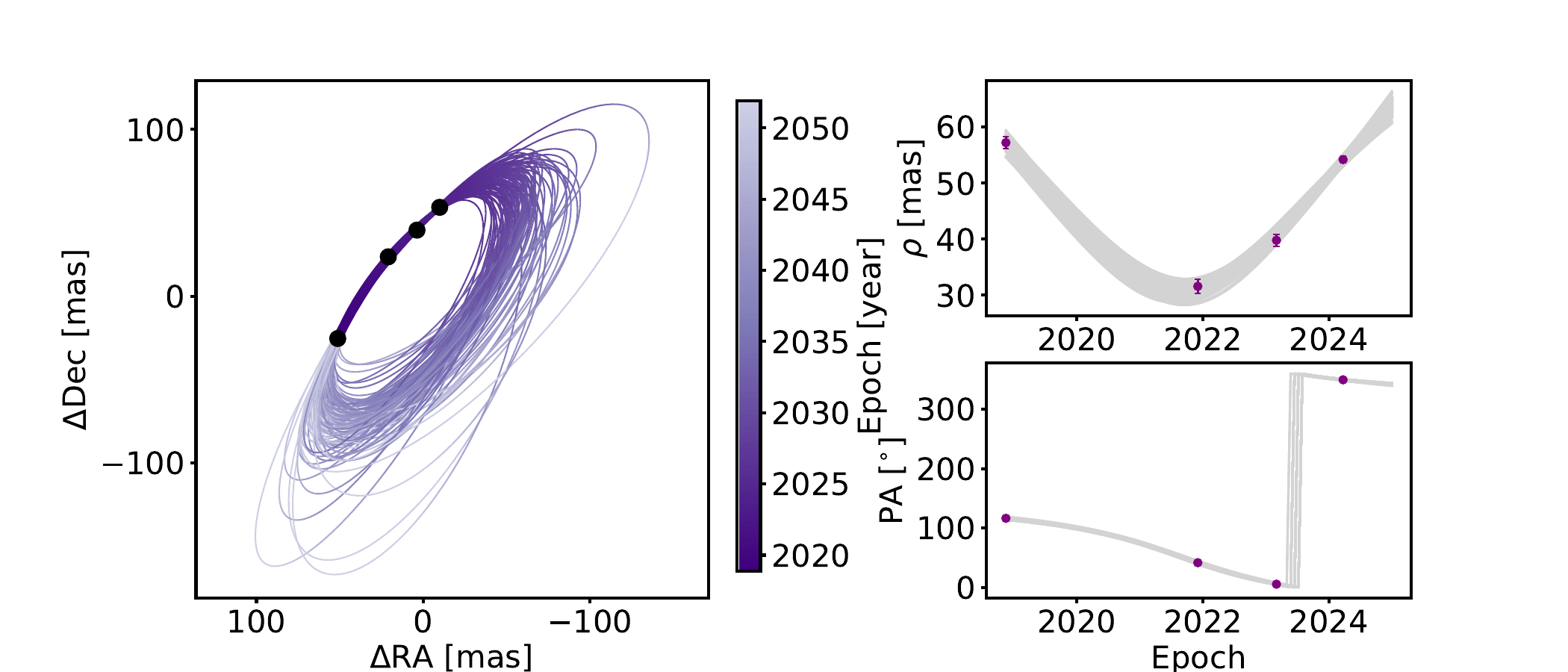}  
\caption{100 orbits drawn from the posterior distributions of the binary fitted with \textsc{orbitize}.
} 
\label{fig:binary_orbitize_orbit}
\end{figure*}

We used \textsc{orbitize} \citep{Blunt2020} to find any plausible orbital solutions for the two stars by fitting the positions measured from direct imaging observations.
Because the observations in $K$s, $K$12, and $J$23 were taken very close to the observations in the $H$ band, we only fitted the positions in the three $H$-band epochs and the last $z\prime$-band epoch.
We set a Gaussian prior for the parallax with 7.3\,mas reported in the Gaia catalogue and inflated its uncertainty to 0.5\,mas. We set a broad Gaussian prior for the total mass with 1.8 $\pm$ 0.8\,\msun\ with an initial guess from the flux ratio.
We show the posterior distributions of orbital parameters in Fig.~\ref{fig:binary_orbitize_corner} and 100 randomly drawn orbits in Fig.~\ref{fig:binary_orbitize_orbit}.
A high inclination of 110\fdg4 $\pm$ 3\fdg1 is required to fit the orbit.
The fitted total mass is 2.1 $\pm$ 0.5\,\msun{}.
The mass of the primary star is well-constrained to about 1.03\,\msun{} (see Sect~\ref{sec:age}).
This implies that the best-fitted mass of the fainter star is 1.06\,\msun{}, resulting in a mass ratio between the fainter star and brighter star of 1.
However, a mass around 1\,\msun{} for the fainter star contradicts the mass expected from the measured flux ratio between the two stars. From the measured flux ratio, we expect the fainter star to be considerably less massive than the brighter star.

Furthermore, using \textsc{gaiaunlimited} \citep{Castro-Ginard2024} which queries Gaia scanning laws and estimates Gaia DR3 detectability of unresolved binary systems, we calculated a RUWE of 1.0 -- 2.4 for the orbits fitted by \textsc{orbitize} depending on the initial phase of the orbits observed by Gaia, much smaller than 6.012.
We also tried to fit the orbit of the fainter star with \textsc{Stan} \citep{Wallace2025} combining RUWE and the direct imaging observations. \textsc{Stan} is developed to fit the mass and orbit of a hidden companion that can cause a high RUWE of a given star by fitting Gaia absolute astrometry including proper motion, parallax and RUWE and possibly combined with image observations using Bayesian inference.
However, it is not able to find plausible orbit solutions for the two stars.
While we can constrain the orbit very well by including RUWE and positions only between 2018 and 2023 in the orbital fitting, the observed position in 2024 does not agree with the predicted position of 2024, which should show an arc instead of still linear motion.
It is therefore also problematic to explain the high RUWE value given these orbital solutions for the two stars.
We calculated the PMA from the orbital fit posteriors following the method of \citet{Biller2022}, adopting masses for the fainter star of 0.1, 0.3, 0.5, 0.7, and 0.9\,\msun{}. Only secondary masses $>$0.7\,\msun{} can produce PMA on the order of the measured PMA for this system, however, such a high mass for the secondary is in tension with the measured flux ratio between the two stars.

\subsection{SED fitting}
\label{sec:sed}
We performed spectral energy distribution (SED) fitting with \textsc{species} \citep{Stolker2020b} to the two stars simultaneously using the multimodal nested sampling algorithm \citep{Feroz2009}.
We fitted broadband photometry (unresolved) of 2M1006 from GAIA, 2MASS and WISE catalogues with the synthetic spectra of BT-Settl-CIFIST \citep{Allard2013}.
We used the parallax from Gaia with an inflated error of 0.5\,mas.
We set uniform priors for the \teff{}, log(g), and radii of both stars: 4000\,K $<$\teff{}$_1 <$ 7000\,K, 3000\,K $<$ \teff{}$_2 <$ 4000\,K, 3\,dex $< \log(g_1) <$ 6\,dex, 3\,dex $< \log(g_2) <$ 6\,dex, 0.005\,\rsun\ $< R_1 <$ 2\,\rsun\, and 0.005\,\rsun\ $< R_2 <$ 2\,\rsun{}, and solar metallicity.
To combine our direct image observations with broadband photometry, we set Gaussian priors for the flux ratio of the fainter star to the primary star in SPHERE/IRDIS/B\_$H$, SPHERE/IRDIS/B\_$K_S$, SPHERE/IRDIS/D\_$J$23 and MagAO-X/camsci1/$z'$ bands.
The optimised SED fitting is presented in Fig.~\ref{fig:sed_star}.
The posterior distributions of the fitted parameters are presented in Appendix~\ref{app:sed_post}.

We derived \teff$ = 5196^{+162}_{-96}$\,K, R$ = 1.09 \pm 0.05\,\rsun{}$, $\log(g) = 3.88^{+0.61}_{-0.50}$\,dex, $\log(L/L_{bol}) = -0.10 \pm 0.05$\,dex for the primary star and \teff$ = 3458^{+110}_{-91}$\,K, $R = 0.99 \pm 0.06\,$\rsun{}, log($g$) = 4.48$^{+0.59}_{-0.43}$\,dex, log($L/L_{bol}$) = -0.90 $\pm$ 0.04\,dex for the fainter star. 
With an estimation of \teff{} for young stars from \cite{Pecaut2013}, the primary star should be a G8--K0 star.
The radius of the primary star is constant with its \teff{} for a 20 -- 30\,Myr 1\,\Msun{}, star from the BT-Settl evolutionary model \citep{Baraffe2015}. 
However, the radius of the fainter star is too large for an M dwarf with that \teff{}.
It has to be younger than 4\,Myr to have a radius of 0.99\,\rsun{} from the evolutionary models, but it should have a radius smaller than $\sim$0.7\,\rsun{} if it is in the same system as the primary star with the same age.
There is likely some additional luminosity from a third unresolved object and the inflated radius of the fainter star in the SED fitting is accounting for that additional luminosity.
The fitted $A_V$ is smaller than 0.13\,mag.
No infrared excess is detected from the SED fitting.

\begin{figure}
\centering
  \includegraphics[width=0.5\textwidth]{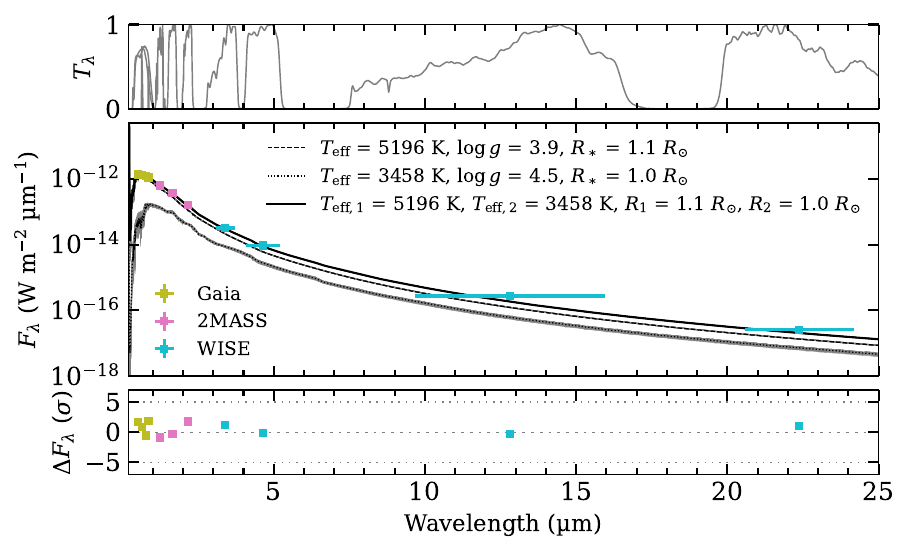} 
\caption{Simultaneous SED fitting of the primary star and fainter star to the unresolved broadband photometric measurements of Gaia, 2MASS and WISE with \textsc{species}.
The black solid line is the combined best-fitted spectrum of the two stars. The black dashed line is the best-fitted spectrum of the primary star, and the black dotted line is the best-fitted spectrum of the fainter star.
Spectra drawn from 30 random samples of each star are shown in grey lines.
The top panel shows the filter transmission profiles and the bottom panel shows the fitting residuals.
} 
\label{fig:sed_star}
\end{figure}

\subsection{Radial velocity}
\label{sec:radial_velocity}
\begin{figure*}
\centering
  \includegraphics[width=0.9\textwidth]{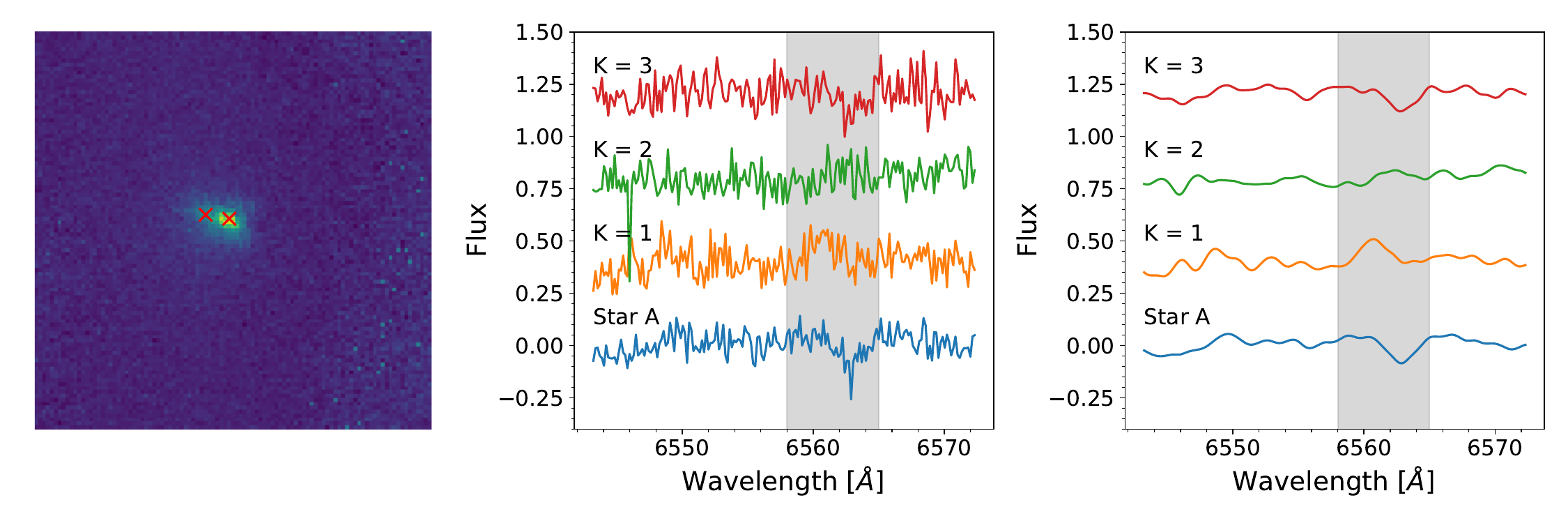} 
\caption{MagAO-X/IFU data of the binary.
Left panel: IFU images of the binary stacked along the wavelength channel.
The image is not rotated to place North up.
Middle panel: star A's spectrum and the first three principal components of star B's spectra. The grey area shows the H$_\alpha$ line band.
Star A presents H$_\alpha$ absorption.
The first component presents H$_\alpha$ emission from star B.
The third component (K=3) is the contamination from star A.
Right panel: smoothed version of the middle panel by a Gaussian kernel with a standard deviation of 0.48\,\AA{}. The grey area shows the H$_\alpha$ line band.}
\label{fig:spectra_ifu}
\end{figure*}

The primary star is detected at high significance while the fainter star is detected with a lower SNR in the MagAO-X/VIS-X IFU data as shown in the left panel of Fig.~\ref{fig:spectra_ifu}.
The spectrum of the primary star is extracted with an aperture of a radius of 3.5 spaxels centred on the star.
For the fainter star, we extracted its spectra with the same aperture but shifted the aperture around the fainter star by 1 spaxel in either x or y direction each time. Starting from the position of the fainter star, x and y change between [-3, 3], respectively, with a total of 49 centring positions.
This results in 49 spectra of the fainter star extracted from different places.
This step is to include as many pixels containing the signal of the fainter star as possible while minimising the effect of possible bad pixels such as those affected by cosmic rays.
Then we applied PCA to the 49 spectra to extract the spectra of the fainter star and also remove the contamination of the primary star.
We compare the spectra of the primary star and the first three components of the 49 spectra in the middle panel of Fig.~\ref{fig:spectra_ifu}.
We detected H$_\alpha$ absorption for the primary star, in agreement with the literature.
We extracted the spectrum of the fainter star as the first principal component as the fainter star is the dominant source in these regions.
We detected H$_\alpha$ emission in the first component, suggesting that the fainter star is likely a magnetically active M dwarf.
The second principal component is featureless and likely originates from the background.
The third principal component has the same shape as the primary star's spectrum with the same H$_\alpha$ absorption feature.
This verifies that this method successfully distinguishes the spectra of the two stars.
We also injected artificial signals of the same brightness and blue shift as the fainter star at the opposite side at the same separation to the brighter star. We successfully retrieved the injected signal using the same method, verifying the reliability of this technique. The injection test is demonstrated in Appendix.~\ref{app:rvinj}.

The right panel of Fig.~\ref{fig:spectra_ifu} presents the smoothed spectra convolved with a Gaussian kernel of a standard deviation = 0.48\,\AA{} from one exposure.
The H$_\alpha$ absorption and emission due to the two stars are easy to identify.
We have three exposures and applied the same method to them.
All of them present consistent H$_\alpha$ absorption in the spectrum of the primary star and H$_\alpha$ emission in the first principal component of the fainter star's spectra.
We fitted a Gaussian profile to the spectra and measured a blue shift of 2.11 $\pm$ 0.14\,\AA{} between the H$_\alpha$ emission line centre and the H$_\alpha$ absorption line centre.
This corresponds to a radial velocity difference of $96.5 \pm 6.6$ \kms{}.
The escape velocity of the two stars is calculated by:
\begin{equation}
v_{esc} = \sqrt{\frac{2G (M_1 + M_2)}{r}}
\label{eq:escape}
\end{equation}
The total mass derived from orbital fitting is 2.1 $\pm$ 0.5\,\msun{} and the closest projected separation of the two stars is 32 $\pm$ 1\,mas which is $4.2 \pm 0.1$\,au for the parallax of 7.32 $\pm$ 0.07\,mas.
Putting them into Eq.~\ref{eq:escape}, the upper limit of the escape velocity is $29.6 \pm3.6$ \kms{}.
The radial velocity difference between the two stars measured in the H$_\alpha$ line is much larger than their escape velocity.

We are cautious about this comparison because H$_\alpha$ emission profiles of M dwarfs can vary due to chromospheric activity, although it is rare to produce such a large blue shift by chromospheric activity.
An analysis of H$_\alpha$ emission lines of 72 M dwarfs finds that 94\% of the stars have blue shifts smaller than 50 \kms{} at H$_\alpha$ emission lines \citep{Flasseur2018}.
We measure a full width at half maximum (FWHM) of 105 \kms{} of the H$_\alpha$ emission line for the fainter star.
But all stars that have such large FWHM in \cite{Flasseur2018} have a shift $<-50$ \kms{} and more than half of them show red shifts.
Therefore, while we could not ascertain that the two stars are not gravitationally bound from radial velocity measurements, it makes this two-body system more debatable.

\subsection{A third companion?\label{sec:triplesystem?}}
There are tensions in the data treating the two stars as a binary system: 
\begin{enumerate}
    \item The dynamical mass measured for the fainter star is too high to match the observed flux ratio between the two stars;
    \item The fitted orbit from direct imaging does not agree with the acceleration measurement for this system; 
    \item The radius of the primary star from the SED fitting is too large if it has the same age as the primary star;
    \item The radial velocity difference between the two stars is too large for them to remain gravitationally bound.
\end{enumerate}

We used the equation from \cite{Brandner2000} to estimate the probability that the fainter star is a randomly aligned foreground or background star:
\begin{equation}
    \centering
    P(\Theta, m) = 1 - e^{-\pi \rho(m) \Theta^2},
\end{equation}
where $\Theta$ is the angular separation of the object to the central star, $\rho(m)$ is the cumulative surface density of background sources of limiting magnitude $m$. We queried all sources within 10 arcmin of the primary star from the 2MASS catalogue \citep{Skrutskie2006} and counted stars with magnitude down to 1 mag fainter than the faint star, which is 67 stars. Then we estimate the probability of the two stars being unrelated is $5.22 \times 10^{-7}$.

Given that the probability of the two stars being unrelated is extremely low, we hypothesise that there is at least one additional unseen companion in the system that may explain the disagreements in our current analysis.
Either the primary star or the fainter star can have an additional unresolved stellar component.
If it is related to the brighter star, this could explain the PMA, RUWE and SED and also bias the measured $\Delta$RV. 
If it is related to the fainter star, this can increase the total mass of the fainter star.
For example, if B is an equal mass binary, then all other SED parameters remain the same but the radius of each object decreases from 1.0\,\rsun{} to 0.7\,\rsun{} and the individual masses are $\sim$0.4\,\msun.
Then the total mass of the fainter component is 0.8\,\msun.
The mass ratio is 0.8 and meets the required total mass of orbital fitting. 
While the PMA and $\Delta$RV may also be explained by the multiplicity of the fainter star, the RUWE of A cannot be explained by the derived acceleration: with a mass ratio of 1, the derived highest RUWE is only 2.4.

Besides the astrometric perturbations caused by a companion, there are other means of producing a RUWE excess, such as variability \citep{Belokurov2020}.
In TESS lightcurves of the system, the primary is variable at a level of $\sim 2$\% with a period of 0.73\,d.
There are two flare events and one of them increases the total brightness by 4\% and another very strong eruption event at the end of the lightcurve which increases the total brightness by 10\%.
These flare events are irregular in period and intensity and are consistent with flaring events seen from low mass stars, which we attribute to coming from B.
A few percentages of variability can lead to a photocentre shift of several mas, which can cause a large RUWE excess.
Therefore, this additional component can either be related to the brighter star or the fainter star.

\subsection{Mass and age estimation}
\label{sec:age}
\begin{figure*}
\centering
  \includegraphics[width=0.4\textwidth]{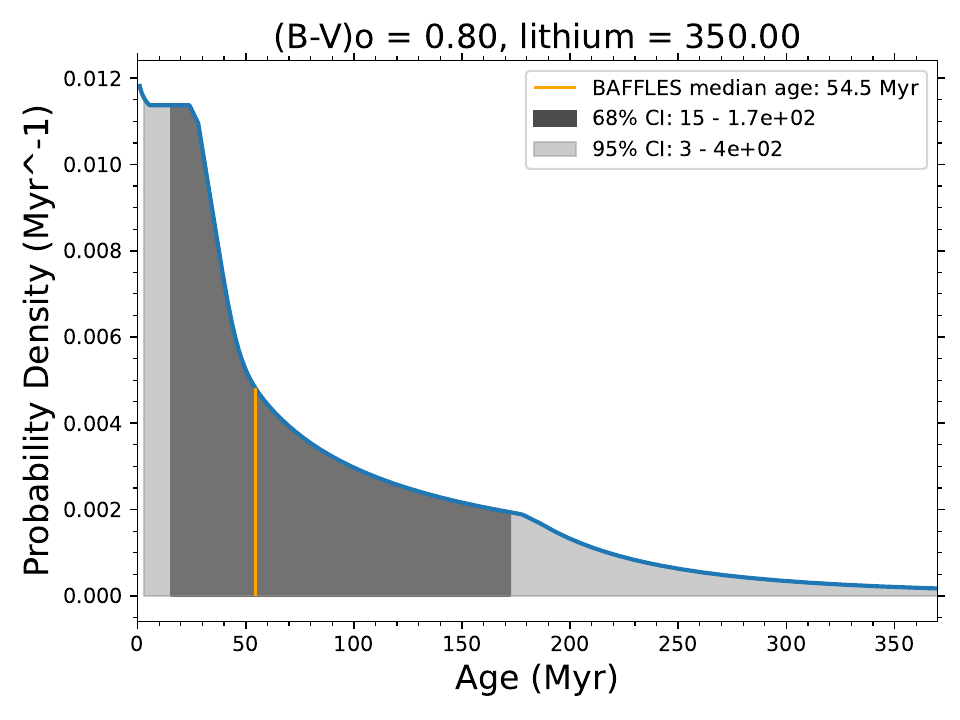} 
  \includegraphics[width=0.4\textwidth]{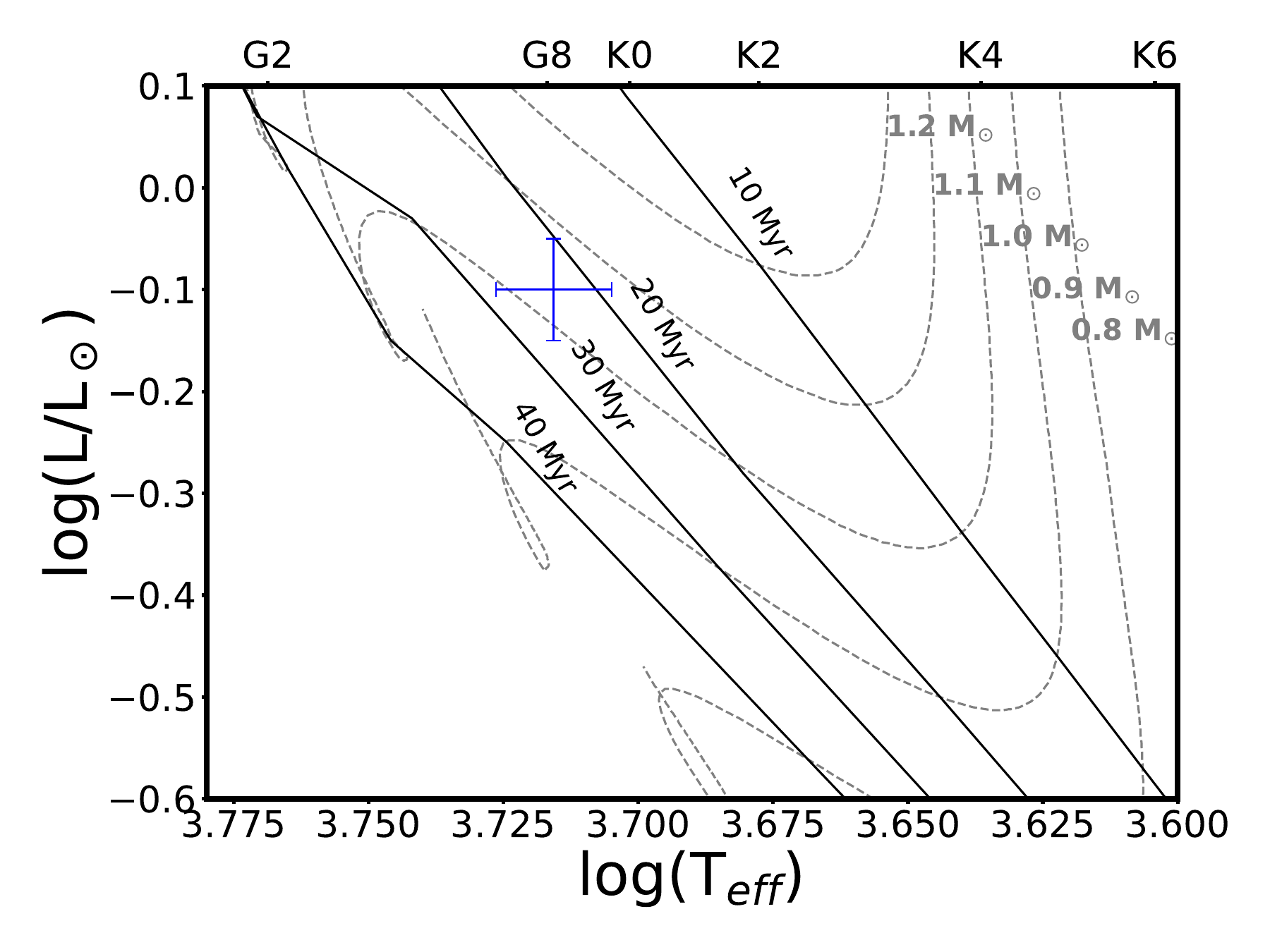} 
\caption{Age estimation of 2M1006.
Left panel: age estimation by Li depletion using BAFFLES \citep{Stanford-Moore2020}.
Right panel: age estimation by combining the luminosity and \teff{} of the primary star from SED fitting with the isochrones and tracks of BT-Settl evolutionary model \citep{Baraffe2015}.
} 
\label{fig:yses3age}
\end{figure*}

The unresolved star is variable with a $V$-band amplitude of $\sim$0.07 mag and rotation period $P_{rot}$ = 0.7271\,d \citep{Kiraga2012}. 
Adopting the colour conversions from \cite{Riello2021} and the Gaia DR3 $G$ and $BP-RP$ colour, we estimated $V$ = 11.015\,mag, which is in the middle of the previously published estimates.
We used two methods to estimate the age of the star.
The first one is to estimate the age by Lithium depletion.
We fitted the equivalent width of Lithium absorption (EW(Li)) using BAFFLES, Bayesian Ages for Field Lower-Mass Stars \citep{Stanford-Moore2020}. With EW(Li) = 350\,m\AA \citep{Torres2006} and $B-V$ = 0.797\,mag, BAFFLES estimates a median age of 54.5\,Myr with a 95\% confidence range of 2.95 -- 396\,Myr, shown in the left panel of Fig.~\ref{fig:yses3age}. 
The second method is to place the primary star in the luminosity and \teff{} diagram with isochrones and tracks from BT-Settl evolutionary model \citep{Baraffe2015}.
With \logl\,= -0.10 $\pm$ 0.05\,dex and $\log$\,\teff\, = 5196$^{+162}_{-96}$\,K, we estimated a mass of 1.03\,\Msun\, with a 68\% confidence range of 0.99 -- 1.08\,\Msun\, and an age of 25\,Myr with a 68\% confidence range of 19 -- 28\,Myr, shown in the middle panel of Fig.~\ref{fig:yses3age}.
All the results suggest that the primary star is a young star. We adopted the age results from the second method.
The unresolved photometric observations of the system from the literature and adopted parameters of the primary and fainter stars are presented in Table~\ref{tab:yses3_phot}.

\begin{table}[ht]
\begin{center}
\caption{2M1006 Stellar Parameters}
\begin{tabular}{lll}
\hline \hline
Parameter & Value & Ref.\\
\hline
\multicolumn{3}{c}{Unresolved stellar binary}\\
\hline
RA(J2000) & 10:06:55.72 & Simbad \\
Dec(J2000) & -63:52:08.61 & Simbad \\
$B$      & 11.812  $\pm$ 0.014  & APASS/DR9\\
$V$      & 11.015  & this work\\
$g'$     & 11.368  $\pm$ 0.054  & APASS/DR9\\
$r'$     & 10.690  $\pm$ 0.060  & APASS/DR9\\
$i'$     & 10.385  $\pm$ 0.033  & APASS/DR9\\
$G_{Bp}$ & 11.227  $\pm$ 0.008  & GaiaEDR3\\
$G$      & 10.744  $\pm$ 0.004  & GaiaEDR3\\
$G_{Rp}$ & 10.085  $\pm$ 0.006  & GaiaEDR3\\
$G_{Bp}$ - $G$  &    0.482880  & GaiaEDR3\\
$G_{Bp}$ - $G_{Rp}$ & 1.141988 & GaiaEDR3\\
$G$ - $G_{Rp}$   &   0.659108 & GaiaEDR3\\
$J$   & 9.262 $\pm$ 0.028 & 2MASS\\
$H$   & 8.744 $\pm$ 0.061 & 2MASS\\
$K_s$ & 8.580 $\pm$ 0.024 & 2MASS\\
$W1$  & 8.530 $\pm$ 0.023 & allWISE\\
$W2$  & 8.541 $\pm$ 0.020 & allWISE\\
$W3$  & 8.455 $\pm$ 0.018 & allWISE\\
$W4$  & 8.236 $\pm$ 0.149 & allWISE\\
EW(Li\,I\,$\lambda$6707) (m\AA) & 350 $\pm$ 10 & (1)\\
EW(H$\alpha$) (\AA) & 0.00 $\pm$ 0.01 & (1)\\ 
\vsini\ (\kms)  & 77.0  $\pm$ 7.7  & (1)\\ 
$A_V$ (mag) & $<$ 0.13 & this work\\
\hline
\multicolumn{3}{c}{Primary (A)}\\
\hline
SpT & G8V -- K0V & (1) and this work\\
\teff\ (K) & 5196$^{+162}_{-96}$ & this work\\
\logl\ (dex)   & -0.1 $\pm$ 0.05 & this work\\
Age (Myr)    & $25^{+3}_{-6}$ & this work\\
Mass (\Msun)   & 1.03 $\pm$ 0.05 & this work\\
$G$ & 10.82 $\pm$ 0.01 & this work\\
$MagAO-X/z^\prime$ & 9.94 $\pm$ 0.01 & this work\\
$SPHERE/J$2 & 9.56 $\pm$ 0.01 & this work\\
$SPHERE/J$3 & 9.40 $\pm$ 0.01 & this work\\
$SPHERE/H$ & 9.05 $\pm$ 0.01 & this work\\
$SPHERE/K_S$ & 8.96 $\pm$ 0.02 & this work\\
\hline
\multicolumn{3}{c}{Fainter star (B)}\\
\hline
\teff\ (K) & 3458$^{+110}_{-91}$ & this work\\
\logl\ (dex)   & -0.90 $\pm$ 0.04 & this work\\
$G$ & 13.67 $\pm$ 0.13 & this work\\
$MagAO-X/z^\prime$ & 11.80 $\pm$ 0.06 & this work\\
$SPHERE/J$2 & 11.04 $\pm$ 0.05 & this work\\
$SPHERE/J$3 & 10.83 $\pm$ 0.04 & this work\\
$SPHERE/H$ & 10.30 $\pm$ 0.04 & this work\\
$SPHERE/K_S$ & 10.04 $\pm$ 0.03 & this work\\
\hline
\end{tabular}
\label{tab:yses3_phot}
\tablebib{(1)~\citet{Torres2006}.
}
\tablefoot{The magnitudes of the primary star and fainter star between $G$ and $SPHERE/K_S$ are calculated by synthetic photometry on the fitted spectra with \textsc{species}. Their uncertainties are likely underestimated.}
\end{center}
\end{table}

\section{The planetary-mass candidate}
\label{sec:astrophot}
\begin{figure}
\center
\includegraphics[width=0.4\textwidth]{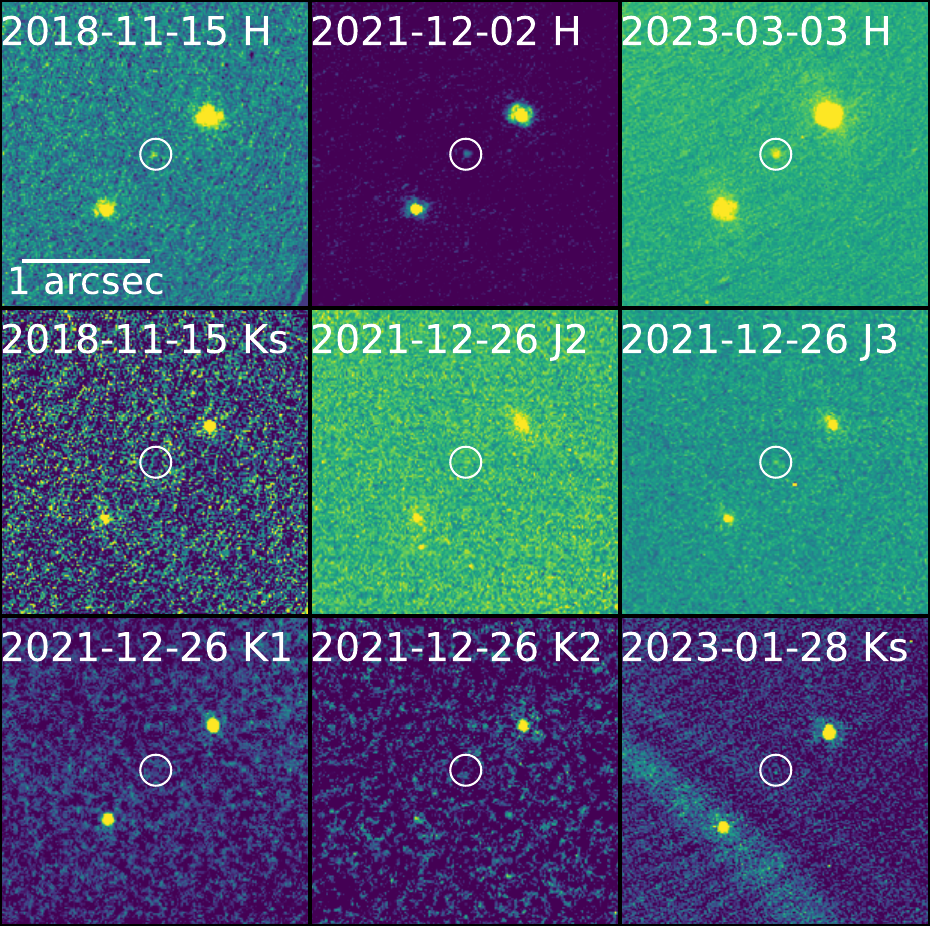} 
\caption{Images of the candidate in five epochs of $H$, $J$23, $K$12, and $K_S$ bands.
The candidate is the faint source between the two bright background stars highlighted by the white circles.
It is only visible in the $H$ band with marginal detection in $J$23 bands.
} 
\label{fig:yses3b_all}
\end{figure}

Figure~\ref{fig:wholeime} shows all the sources and the central binary in the $H$ band.
The deepest images are in the $H$ band, with ten sources in the field of view (FoV).
Several sources are not detected in the other bands, including the candidate.
We detected the candidate in the $H$ band as shown in Fig.~\ref{fig:yses3b_all} with marginal detection in $J$23 bands.

\subsection{Photometric analysis}
We performed aperture photometry to measure the flux of the sources in the FoV and PSF fitting to measure the flux of the central two stars.
We used the same aperture with a radius equal to the mean FWHM of all sources.
For each source, we took the three $\sigma$ clipped median of an annulus centred on the source with radii ranging from 10 to 15 pixels as the background level per pixel and subtracted it from the aperture photometry.

To measure the signal-to-noise ratio (S/N), we chose 14 random apertures of the same size as the aperture on the source adjacent to the source and took their standard deviation as the noise level.
Table~\ref{tab:yses3b_phot} summarises the contrast of the candidate to the primary star and S/N.
We consistently detect the candidate in the $H$ band with S/N $>$ 5 but with only upper detection limits in $K$12 and $K_S$ bands and marginal detection in $J$23 bands.
We achieved the highest detection significance on 2023-03-03 with S/N = 8.8.
Averaging the magnitude of the candidate in three measurements, its apparent magnitude in the $H$ band is 22.04 $\pm$ 0.13 mag.

\begin{table}[ht]
\begin{center}
\caption{Photometric measurements of the candidate.}
\begin{tabular}{llll}
\hline \hline
Date & Band & $\Delta$Mag (mag) & Detection S/N \\
\hline
2018-11-15 & $H$ & 12.89 $\pm$ 0.21 & 5.2\\
2021-12-02 & $H$ & 12.93 $\pm$ 0.19 & 5.6\\
2023-03-03 & $H$ & 13.15 $\pm$ 0.12 & 8.8 \\
2018-11-15 & $K_S$ & $>$ 11.24 & -1.3 \\
2021-12-26 & $J2$ & 13.52 $\pm$ 0.44 & 2.5 \\
2021-12-26 & $J3$ & 12.50 $\pm$ 0.25 & 4.4 \\
2021-12-26 & $K1$ & $>$ 12.55 & 1.0 \\
2021-12-26 & $K2$ & $>$ 11.34 & -0.5 \\
2023-01-28 & $K_S$ & $>$ 12.68 & -0.7 \\
\hline
\end{tabular}
\tablefoot{We only detect the candidate in the $H$ band with marginal detection in $J23$.
The contrast is relative to the flux of the primary star and is the 5$\sigma$ contrast for bands of no detection.
Detection S/N is calculated by aperture photometry.}
\label{tab:yses3b_phot}
\end{center}
\end{table}

\subsection{Astrometric analysis}
\label{sec:yses3b_astro}
We fitted a 2D Gaussian model to the candidate to measure its position. At first, we fitted a 2D Gaussian model to the candidate with \textsc{Astropy} \citep{Astropy2013,Astropy2018} to obtain the initial guess of the position. Then we cut out a sub-image of 31 pixels $\times$ 31 pixels at the position of the candidate. We took three $\sigma$ clipped median and standard deviation of an annulus with an inner radius of 4 pixels and an outer radius of 12 pixels from the central position of the candidate as the local sky background and uncertainty of each pixel in the sub-image, respectively.
We subtracted the fitted model from the sub-image and minimise residuals by minimising the chi-square $\chi = \Sigma^{N}_{i=1} \frac{(Data - Model)^2}{Data~error^2}$ (N is the number of pixels) using the {\sc emcee} package \citep{ForemanMackey2013}.

The uncertainty from the posterior distribution is tiny, and therefore we injected artificial stars using the PSF template and retrieved positions using the above algorithm. 
We injected 50 stars at the same separation of the object but at the opposite side of the image with one star per injection. We took three $\sigma$ clipped mean of the deviations between the injected and retrieved positions of the 50 injections as the uncertainty of our position measurement algorithm.
We also included uncertainties from pixel scale (12.25 $\pm$ 0.021 mas per pixel), true north angle correction (-1.75 $\pm$ 0.08 deg), pupil offset (135.99 $\pm$ 0.11 deg), distortion (0.4 mas per 1 arcsec separation) and systematics such as coronagraph centring uncertainty (we added arbitrary 1\,mas in RA and Dec) in the propagation of the error when calculating the separation between the primary star and the candidate. 
The error budget is demonstrated in Table~\ref{tab:error_table}.

\begin{figure*}
\centering
  \includegraphics[width=0.4\textwidth]{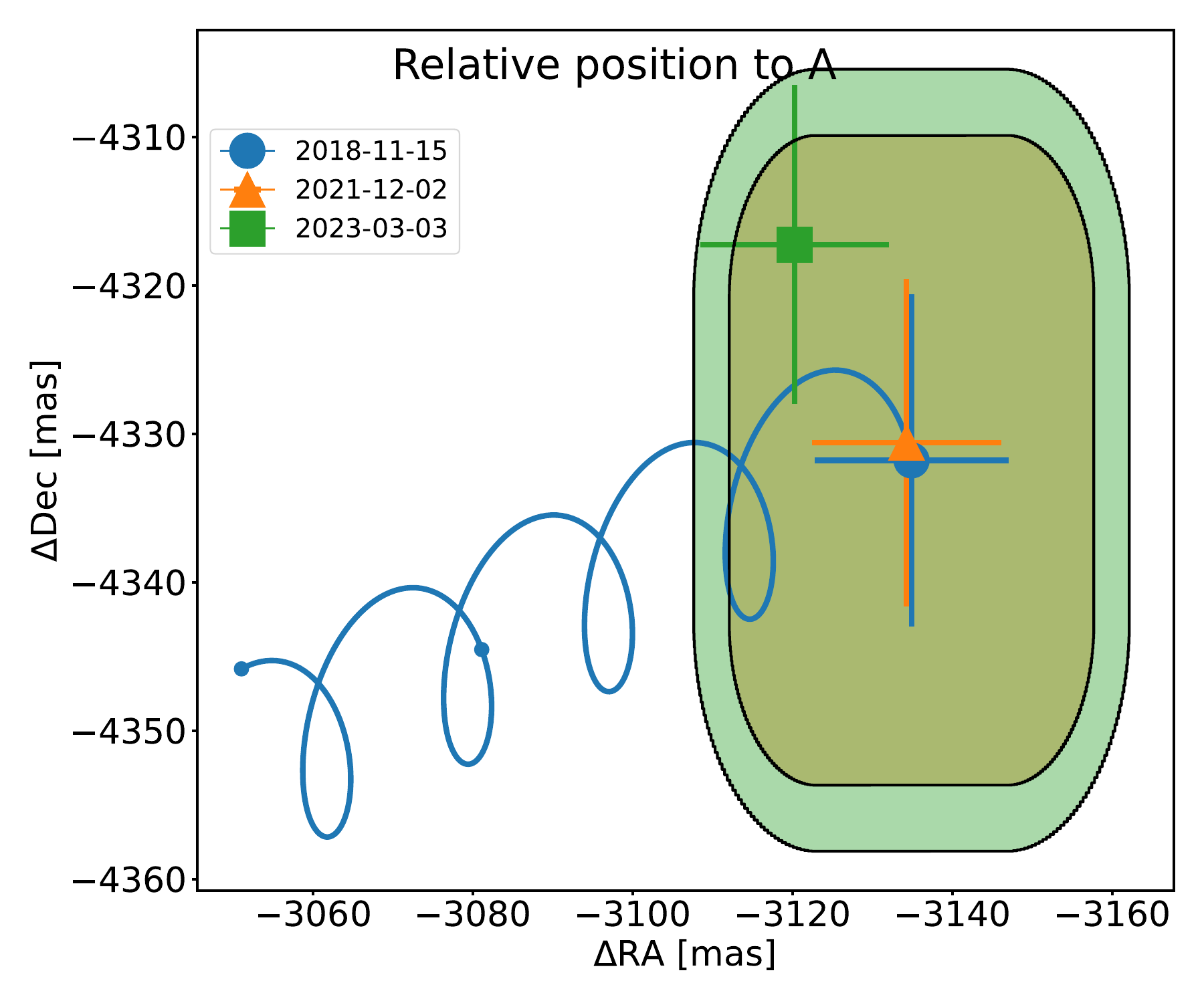} 
  \includegraphics[width=0.4\textwidth]{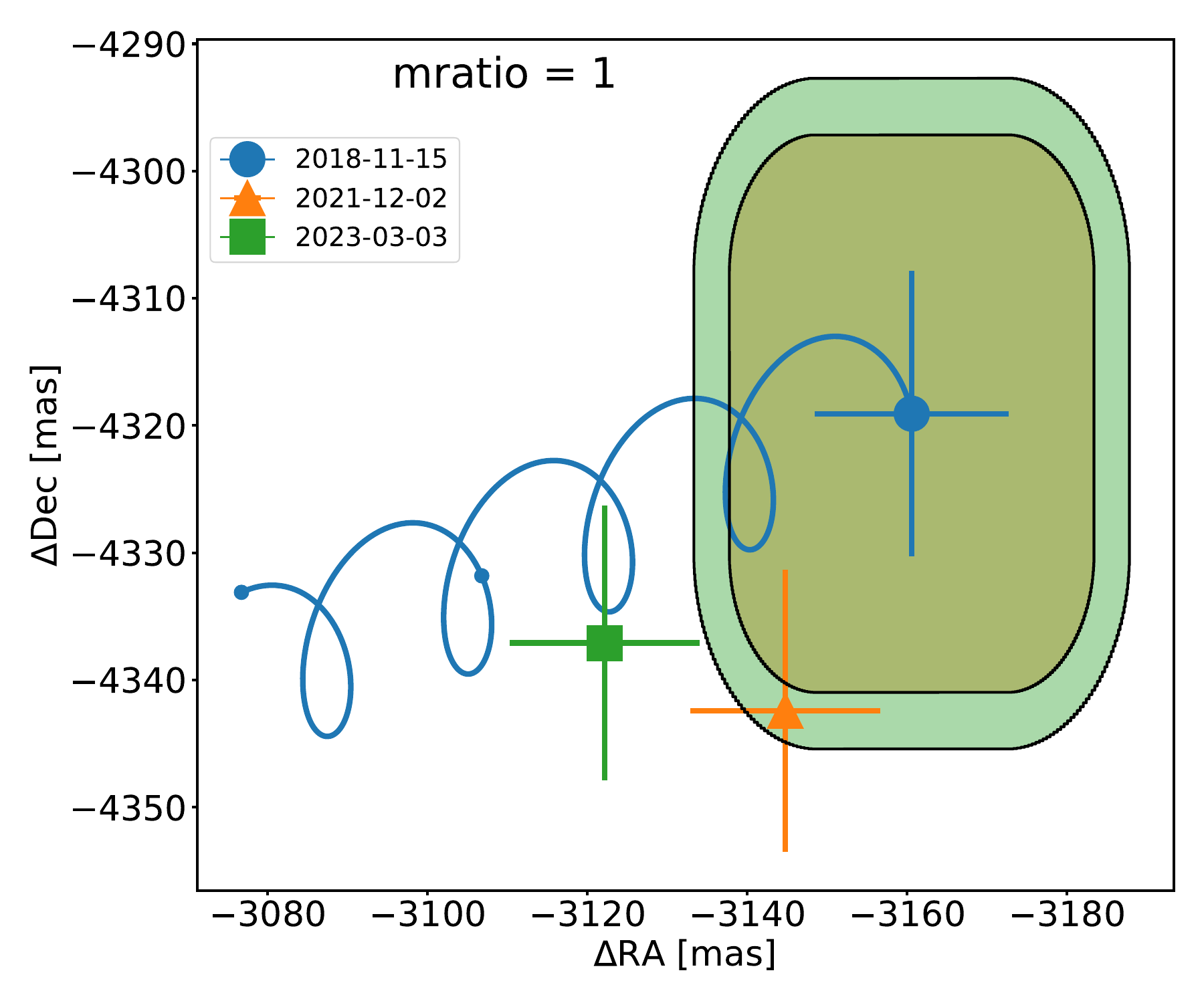} 
\caption{Relative astrometry of the candidate in RA and Dec offsets.
Left panel: relative astrometry of the candidate to the primary star (mass ratio = 0). 
Right panel: relative astrometry of the candidate to the barycentre if the fainter star and primary star are an equal mass binary (mass ratio = 1)
The yellow region and green region are the candidate companion positions allowed by the escape velocity within 1$\sigma$ on 2021-12-02 and 2023-03-03.
The blue line is the trajectory of a static background star evolving from 2018-11-15 calculated with the proper motion and parallax reported in Gaia and the two points on the track are the positions on 2021-12-02 and 2023-03-03.}
\label{fig:astrometry_candi_toA}
\end{figure*}

Table~\ref{tab:yses3b_pos_toA} presents the separations and position angles of the candidate relative to the primary star from 2018 to 2023.
The positions of the candidate relative to the primary star between 2018 and 2023 are shown in the left panel in Fig.~\ref{fig:astrometry_candi_toA}. They match within 1$\sigma$, suggesting that this candidate's proper motion is comparable to the proper motion of the primary star.
Evolving from the position on 2018-11-15, it is more than 3$\sigma$ away from the position of a static background source on 2023-03-03.
The candidate is at a separation of 730 $\pm$ 10\,au to the primary star assuming they are at the same distance.
However, as noted in Section~\ref{sec:triplesystem?}, the primary star is very likely a triple system, thus, the true barycentre of the system is unknown.
Due to the unknown barycentre of the central stars, it is not possible to confirm this candidate as a comoving source by common proper motion analysis. 
We discuss what this source could be in Sect.\ref{sec:discussion}.

\begin{table}[ht]
\begin{center}
\caption{Relative position of the planet candidate to the primary star.}
\begin{tabular}{lll}
\hline \hline
Epoch (Julian years) & Sep (arcsec) & PA (deg) \\
\hline
2018.8715 & 5.3471 $\pm$ 0.0099 & 215.89 $\pm$ 0.14\\
2021.9186 & 5.3458 $\pm$ 0.0097 & 215.90 $\pm$ 0.14\\
2023.1664 & 5.3268 $\pm$ 0.0095 & 215.86 $\pm$ 0.14\\
\hline
\end{tabular}
\label{tab:yses3b_pos_toA}
\end{center}
\end{table}

\subsection{Colour-magnitude diagram}
\begin{figure*}
\centering
  \includegraphics[width=0.45\textwidth]{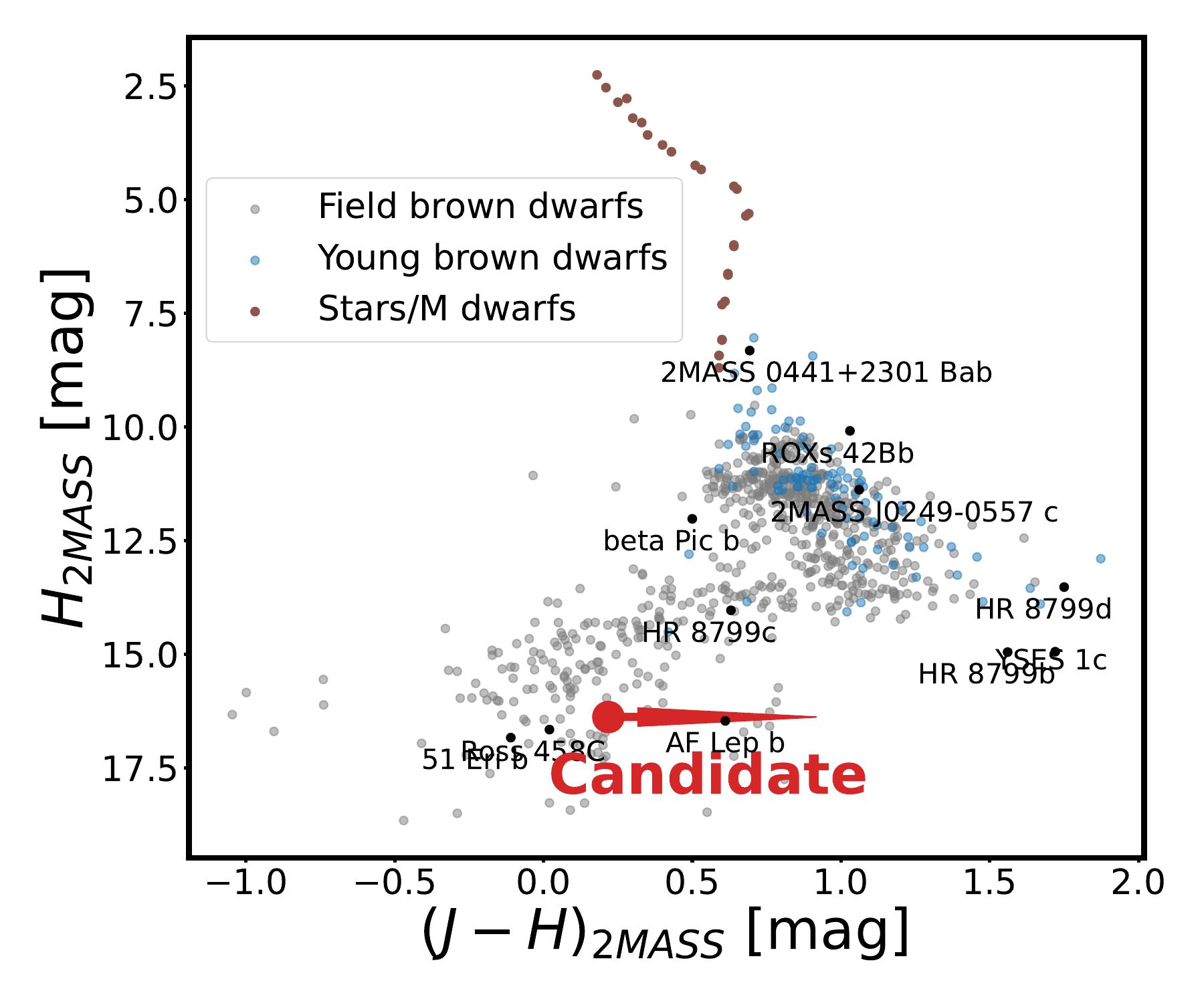} 
  \includegraphics[width=0.45\textwidth]{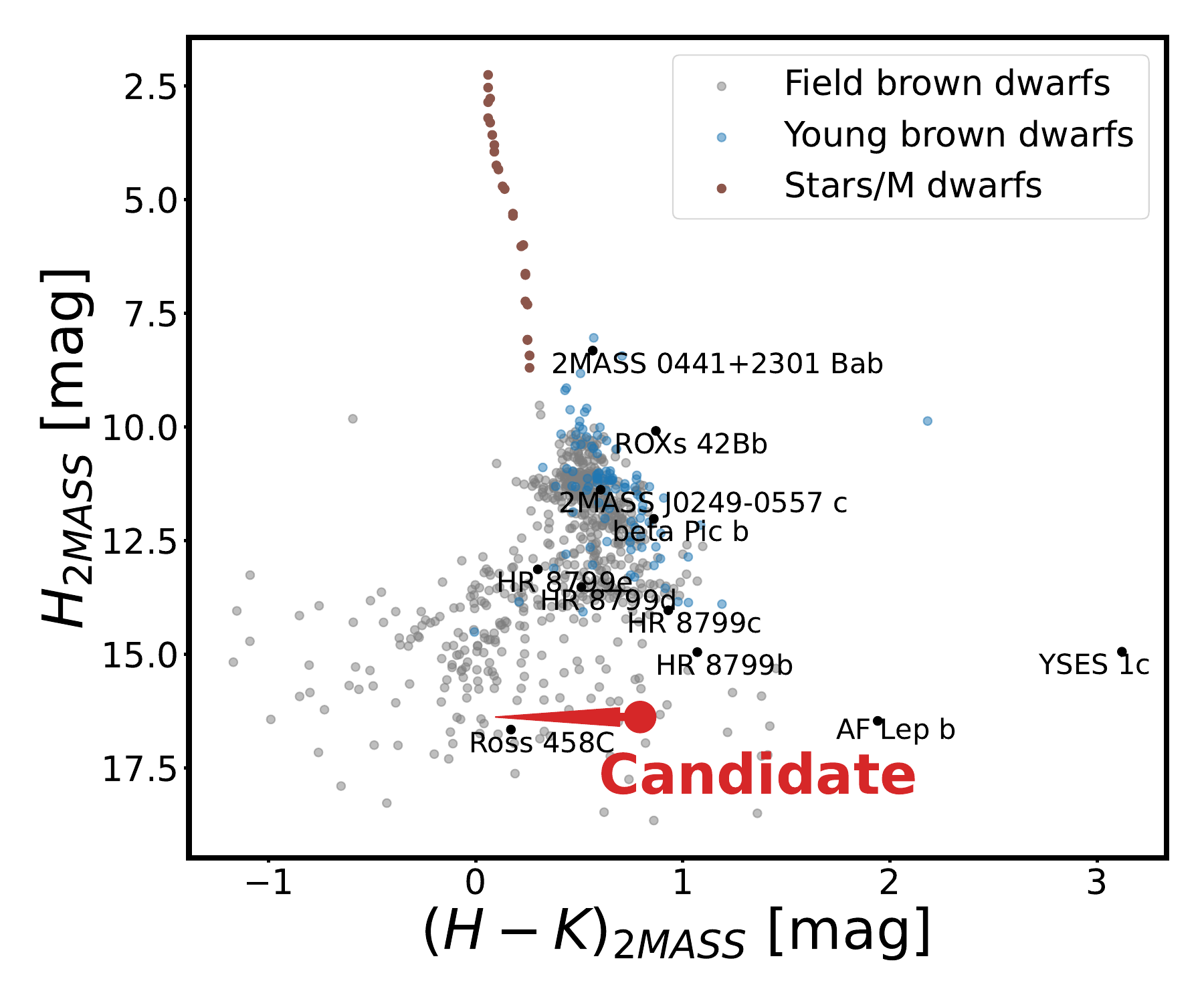} 
\caption{Colour-magnitude diagram of brown dwarfs.
The grey points are field brown dwarfs with high surface gravity. The blue points are young or low surface gravity objects.
The brown points are low-mass stars.
Several directly imaged exoplanets are labelled with black text.
The red arrow shows the possible positions of the candidate assuming it is at the same distance as the central stars.
If confirmed, it would be one of the coolest directly imaged exoplanets.
} 
\label{fig:yses3cmd}
\end{figure*}

Assuming the contrast is the same between $H_{2MASS}$ and $H_{SPHERE}$, we converted the candidate's apparent magnitude to absolute magnitude with the primary star's parallax. 
We placed the candidate in the colour-magnitude diagram (CMD) in Fig.~\ref{fig:yses3cmd}.
Assuming the same distance of the primary star (136\,pc), the candidate has an absolute magnitude of 16.38 $\pm$ 0.13 mag. 
Due to the unknown magnitude in the $J$ band, we could only derive a blue limit of its 2MASS $J-H$ colour of 0.22, redder than most T-type objects of the same assumed $H$ magnitude.
We also present the colour of M dwarfs. 
Based on the current colour range, we cannot determine whether the candidate is a planetary-mass object, brown dwarf, or background M dwarf.
If it is a planet, it would be one of the coolest directly-imaged planets similar to 51~Eri~b \citep{Macintosh2015, Samland2017} and AF~Lep~b \citep{DeRosa2023, Franson2023, Mesa2023}.
Its $J-H$ colour is between these two confirmed planets.
We derive a red limit to its $H-K$ colour of 0.79, redder than most objects of the same $H$ magnitude.
Using the ATMO2020 evolutionary model \citep{Phillips2020}, we estimated a mass of 3--5\,\mj\ under the assumption of chemical equilibrium.

\section{Discussion}
\label{sec:discussion}
\subsection{A comoving planet or free-floating object?}
The discrepancies between the orbital fitting to the central two stars with their RUWE value, SED fitting, and radial velocity measurements suggest that there might be an additional low-mass stellar companion in the system.
The uncertainty in the resulting barycentre of the system makes it challenging to confirm if the candidate is a comoving companion by conventional common proper motion analysis.
This candidate may be a comoving planet, a free-floating planetary-mass object, a brown dwarf or a background star.
We followed the method used in \cite{Nielsen2017} to quantify the probability ratio of the candidate being a background star to a bound planet of the barycentre of the central stars as a function of the mass ratio between the fainter star and the brighter star.
\cite{Nielsen2017} define three likelihood components of each scenario: the overall likelihood of the object being a background star $P(\text{bg})$ or planetary companion $P(\text{pl})$, the relative likelihood as a function of separation from the star $P(\rho | \text{bg})$ or $P(\rho | \text{pl})$, and the relative probability as a function of projected velocity $P(v | \rho_{\text{bg}})$ or $P(v | \rho_{\text{pl}})$:

\begin{equation}
\frac{P_{\text{bg}}}{P_{\text{pl}}} = \frac{P(\text{bg}) P(\rho | \text{bg}) P(v | \rho_{\text{bg}})}{P(\text{pl}) P(\rho | \text{pl}) P(v | \rho_{\text{pl}})},
\label{eq:probility}
\end{equation}

\noindent where $P$ is probability, $\rho$ is projected separation, and $v$ is projected velocity.
Below we describe the calculation taking the mass ratio of 0 as an example, which is an extreme case assuming the barycentre is on the brighter star.
(The positions of the candidate to the barycentre vary with the mass ratio and affect the corresponding terms.)

\begin{figure}
\centering
  \includegraphics[width=0.45\textwidth]{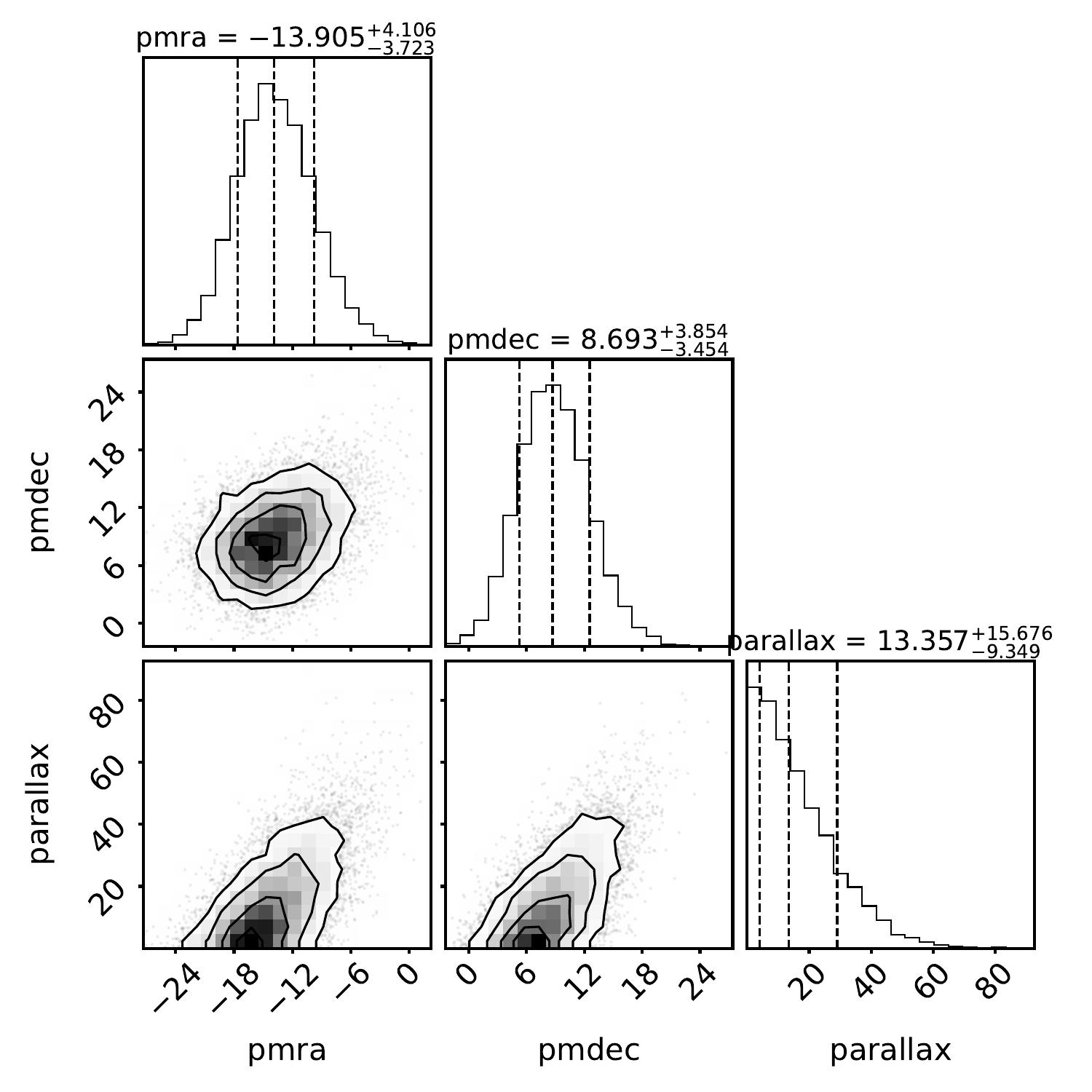} 
\caption{Proper motion and parallax posterior distribution of the candidate assuming it is a free-floating object when the mass ratio of B to A is 0. The proper motion is in the unit of mas yr$^{-1}$ and the parallax is in the unit of mas.
} 
\label{fig:pmpafit_corner_toA}
\end{figure}

\begin{figure}
\centering
  \includegraphics[width=0.5\textwidth]{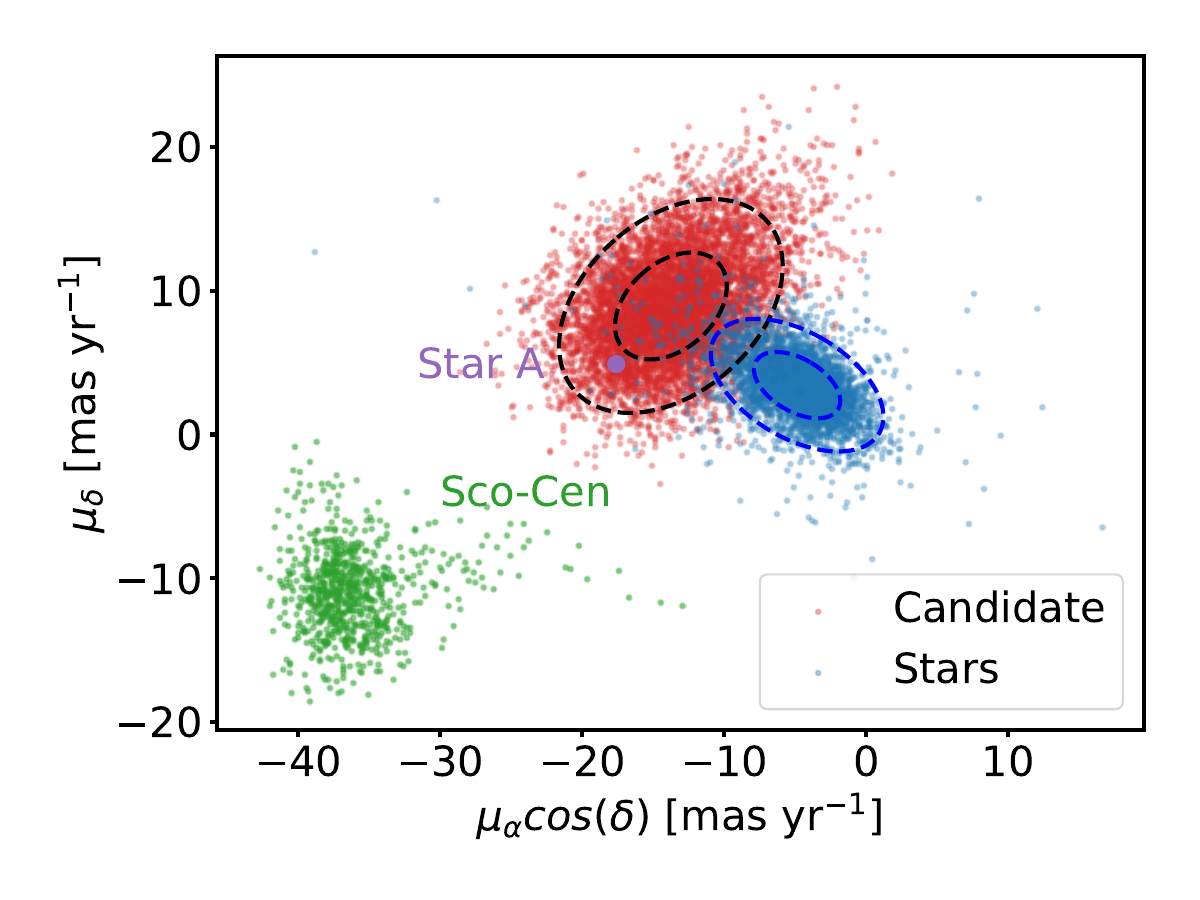} 
\caption{Proper motion of the candidate assuming it is free-floating when the barycentre is on the primary star.
The blue populations are the stars from 0 to 50 kpc within 2$\sigma$ magnitude of the candidate generated by the Besan\c{c}on galaxy model.
The 1 and 2$\sigma$ contours are shown by the dashed blue circles.
The red populations are the posterior distribution of the candidate's proper motion with black circles showing the 1 and 2$\sigma$ range.
The purple point is the proper motion of the primary star reported in Gaia and the errorbar is smaller than the marker size.
The green points are Sco-Cen candidate members within 20 angular degrees of the primary star on the sky.
} 
\label{fig:besancon_pm}
\end{figure}

In the hypothesis that the candidate is a background star, we use the Besan\c{c}on galaxy model \footnote{\url{https://model.obs-besancon.fr/}} \citep{Czekaj2014} to simulate stellar populations in the direction of the stellar system.
We simulated stars with distances from 0 to 50\,kpc within a solid angle of one square degree.
We selected stars with $2MASS$ H magnitudes within 2$\sigma$ of the candidate's magnitude, 21.79 -- 22.30\,mag, which contains 4373 stars.
The SPHERE image covers regions with separations from the bright central star from 0\farcs15 to 5\farcs5, so the first term $P_{\text{b}}$ = 4373 $\times \frac{\pi(5\farcs5)^2 - \pi(0\farcs15)^2}{(3600'')^2}$ = 3.2\%.
Then $P(\rho | \text{b})$ is the ratio of the area of the 2$\sigma$ separation to the whole image, which is $\frac{(5\farcs367)^2 - (5\farcs327)^2}{(5\farcs5)^2 - (0\farcs15)^2}$ = 1.4\%.
For the third term $P(v | \rho_{\text{bg}})$, we assumed the candidate was free-floating and fitted its proper motion and parallax using {\sc emcee}.
We derived its $\mu_\alpha \cos \delta$ = -13.86 $\pm$ 4.02\,mas yr$^{-1}$ and $\mu_\delta$ = 8.70 $\pm$ 3.94\,mas yr$^{-1}$ as presented in Fig.~\ref{fig:pmpafit_corner_toA}.
The parallax is not well constrained because two of the three astrometric measurements are almost exactly one year apart.
We compare its proper motion with the primary star and simulated background stars in Fig.~\ref{fig:besancon_pm}.
The primary star is within the 2$\sigma$ range of the candidate.
Both of them deviate from the majority of the simulated background stars.
Then $P(v | \rho_{\text{bg}})$ is the probability that a background star has a proper motion comparable to the measured proper motion of the candidate. This is calculated by the ratio of the number of background stars that have proper motions within the 2$\sigma$ contour of the candidate to the total number, which is 7.6\%.

In the hypothesis that the candidate is a comoving planet bound to the central stellar system, we started with the companion occurrence rate in the SHINE survey.
\cite{Vigan2021} estimate 5.7\% of FGK stars have companions of 1 -- 75\,\mj\ with a separation range of 5 -- 300\,au.
Due to a lack of observed statistics for planets beyond 300\,au, we used this rate as the lower limit of companion occurrence rate between 5 and 760\,au.
With an age range of 19--28\,Myr and 21.79 -- 22.30\,mag, we estimated a mass range of 2.7 -- 4.5\,\mj\ and used the power-law distribution of planetary mass, $dN / dM \propto M^{-1.31}$ \citep{Cumming2008} to calculate $P(\text{p})$ = 5.7\% $\times$ 0.15 = 0.8\%. 
We used the power-law distribution of separation, $dN / da \propto a^{-0.61}$ \citep{Nielsen2017} and again the 2$\sigma$ range of the separation measured at the first epoch to calculate $P(\rho | \text{p})$ = $\frac{(5\farcs367)^{0.39} - (5\farcs327)^{0.39}}{(5\farcs5)^{0.39} - (0\farcs15)^{0.39}}$ = 0.4\%.
The relative measured projected velocity between the candidate and star A (from 2018 to 2023) is 4.8 $\pm$ 3.8 mas yr$^{-1}$ (3.4 $\pm$ 3.9 mas yr$^{-1}$ in RA and 3.4 $\pm$ 3.6 mas yr$^{-1}$ in Dec), marginally consistent with no relative motion given the measured uncertainties.
The escape velocity for a companion at a similar projected distance is 3.5 $\pm$ 0.5 mas yr$^{-1}$ where the uncertainty is due to the uncertainty in the total stellar mass.
This velocity is the upper limit of the projected escape velocity due to the unknown orbital inclination and phase.
We therefore calculated $P(v | \rho_{\text{pl}})$ by generating 10$^7$ points assuming Gaussian distributions for the measured projected velocity $\sim \mathcal{N}(4.8,3.8^{2})$ and escape velocity $\sim \mathcal{N}(3.5,0.5^{2})$ -the upper limit of the fraction of the points with measured projected velocity smaller than the escape velocity is 36.3\%.
Then,
\begin{equation}
\frac{P_{\text{b}}}{P_{\text{p}}} \sim \frac{3.2\% \times 1.4\% \times 7.6\%}{0.8\% \times 0.4\% \times 36.3\%} = 2.8.
\label{eq:planet_prob_unbound}
\end{equation}

\begin{figure}
\centering
  \includegraphics[width=0.45\textwidth]{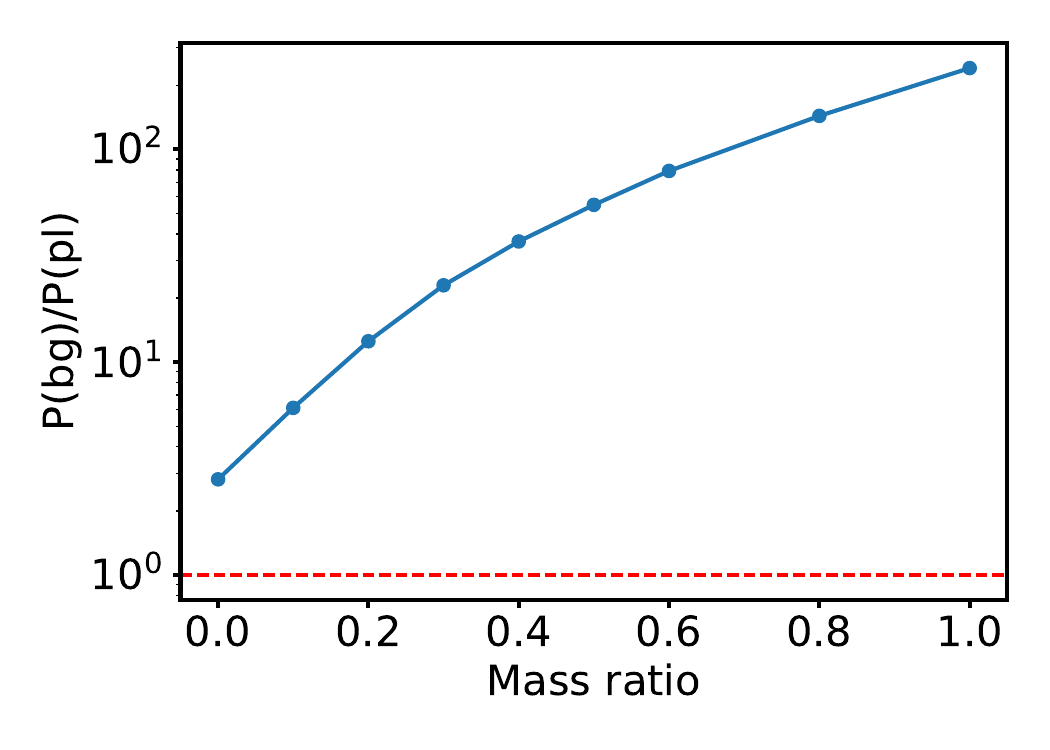} 
\caption{Probability ratio of the candidate being a free-floating object to a bound planet of the central stars as a function of the mass ratio of the fainter star to the brighter star.
The red horizontal line is where the probability ratio = 1.} 
\label{fig:probability}
\end{figure}

The candidate is three times as likely to be an unbound object than a bound planet assuming that the mass ratio of B to A is zero.
We then repeated this calculation increasing the mass ratio to unity and show the probability ratio in Fig.~\ref{fig:probability}.
The reason for this change is mainly due to two factors: the fitted proper motion of the candidate assuming it is free floating, and the measured projected velocity of this candidate assuming it is bound.
It is 3--240 times more likely being a background star than a bound planet with the central star mass ratio varying between 0 and 1.

Though the probability of the candidate being a bound planet is lower compared to that of being an unbound object, we note that there are several confirmed planets beyond 300\,au, such as YSES~1c \citep{Bohn2020c}, b~Cen~b \citep{Janson2021}, and COCONUTS-2b \citep{ZhangZ2021b} and therefore we cannot definitely rule out the possibility of this candidate being a planet by statistical analysis.
Spectroscopic confirmation is therefore crucial to determine if a source is a planetary-mass object, brown dwarf or a background star.

In addition, the statistical analysis is based on multiple assumptions:
\begin{enumerate}
\item Our knowledge of the wide-orbit planet occurrence rate is incomplete: there are no statistical results on the planet occurrence rate beyond 300\,au.
We used two approximations: 
a).the planet occurrence rate is based on separations within 300\,au of 150 stars from multiple stellar regions \citep{Vigan2021} while the planet frequency statistics may vary with regions;
and b).the number of planets as a function of mass is taken from \citet{Cumming2008} which is calculated for radial velocity (RV) detected planets with periods smaller than 2000 days (within 3\,au for solar mass stars). These RV planets are likely formed by core accretion while wide-orbit planets are likely formed by different mechanisms and therefore the planet distribution may not follow this distribution.
\item The proper motion calculation of the candidate - assuming it is a free-floating object - relies on the proper motion of the barycentre, which is approximated by the proper motion of the primary star measured by Gaia.
However, the proper motion of the primary star is likely affected by the fainter star and the potential hidden companion, and the same is true for the proper motion of the barycentre.
\item Due to the unknown orbit of the planet, we adopted the escape velocity as the projected escape velocity, which is the upper limit of the projected escape velocity.
\item We used a galaxy population model to simulate the background star model in the direction of the central star which is a good representative of background star density on a large scale.
However, there is uncertainty about the local background density that is not taken into account. This star is likely located in a crowded field given that there are a total of 10 sources in the FoV besides the central stars.

\end{enumerate}

Given these uncertainties, the probability ratio should not be taken as the deciding factor on the nature of this source.
Regardless, the motion of the candidate is the most significant among the other nine sources in the images and is also significantly distinct from the other sources.
We show the relative motion of all sources in two cases in Fig.~\ref{fig:astrometry_all} in Appendix~\ref{app:relative_astroall}.
If it is a free-floating object, it might be a planetary-mass object, a brown dwarf or a distant background star.
We also show the positions of low-mass stars in Fig.\ref{fig:yses3cmd} using the colour calculated in the evolutionary model of \cite{Baraffe2015} with an age of 1 and 5 Gyr.
Their colour also falls into the colour range of the candidate.
The colous of the candidate are compatible with a wide range of background M dwarfs. For instance, if the candidate was an M dwarf of 0.15\,\msun\ , its absolute magnitude would be 8.7\,mag in $H$.
To match the apparent magnitude of the candidate, it would have to be 4.8\,kpc away and have a very high proper motion within the Galaxy.

\subsection{Group membership}
\label{sec:membership}
The unresolved central star was classified as a member of LCC by \cite{Pecaut2016}.
The age we estimated for the primary star is more than 19--28\,Myr, older than the typical age of LCC members.
LCC is a subregion in Sco-Cen and is revealed to be composed of several substructures of different ages, ranging from 6 to 15 Myr \citep{Ratzenbock2023, Zerjal2023}.
This leads to a question about the membership of the star.
Using the online association membership analysis tool BANYAN $\Sigma$ \citep{Gagne2018}, the star is classified as a field star by 99.9\% with Gaia proper motion and parallax.
Furthermore, we selected Sco-Cen candidate members within 20 angular degrees of the star on the sky from \cite{Luhman2022} and show them in Fig.~\ref{fig:besancon_pm}. The primary star is located outside of the scattering region of these stars and we find that the closest Sco-Cen star to the primary star is at a separation of 8\fdg9. So the primary star is likely not a member of Sco-Cen.
Detailed analysis is required to classify the star's membership, such as including resolved absolute radial velocity measurements instead of using the unresolved radial velocity from Gaia.
This kind of analysis is outside the scope of this work.

\section{Summary}
\label{sec:summary}
We report the detection of a planetary-mass candidate companion imaged around 2M1006.
We consistently detected it in the $H$ band with marginal detection in $J$23 bands and non-detection in $K$12 and $K_S$ bands from 2018 to 2023. 
The central star is resolved to consist of a G8--K0 young Sun analogue with \teff\ = 5196$^{+162}_{-96}$\,K and an M dwarf for the first time. We estimate an age of 19--28\,Myr for the primary star. 
We find tensions between the acceleration data, orbital fitting, SED fitting, and radial velocity difference for the two stars if they are a binary system.
Therefore, we speculate that there is at least one additional low-mass companion in the central stellar system: either the primary star is not a single star or the fainter star is not a single star.
Due to the unknown barycentre of the central stellar system, we are unable to confirm if the candidate is a comoving planet of this system.
Nevertheless, the candidate shows the most significant proper motion compared to other sources in the FoV and shares a common proper motion with the primary star.
Multi-band colours and spectroscopic observations are necessary to identify the properties of this candidate.
Narrow band colour $H$2 - $H$3 can distinguish a low-mass planet with a clear atmosphere from a dusty planet, substellar and low-mass star as the latter three types of objects would have $H$2 - $H$3 $>$0\,mag while a clear-sky low-mass planet would have a very blue $H$2 - $H$3 colour.
Narrow band colour $J$2 - $J$3 can also help identify a clear-sky low-mass planet as it would have $J$2 - $J$3 $>$ 1.5\,mag when the mass is smaller than 10\,\mj{}. 
From the marginal detection in 2021, its $J$2 - $J$3 is 1.02 $\pm$ 0.51\,mag which cannot distinguish it between a low-mass planet, brown dwarf and M dwarf due to the large uncertainty. 
Future high S/N detection in $H$23 and $J$23 may help identity if it is a clear-sky low-mass planet.
Spectroscopic observations can fully confirm its nature.
Due to its faintness, only JWST is capable of obtaining its spectra.
Long-term imaging and radial velocity monitoring of the central stars can ascertain the kinematic nature of the two stars, which is crucial to confirm the companionship of this candidate by common proper motion analysis.
With an apparent magnitude of 22.04 $\pm$ 0.13 mag in $H$ band and the age of the primary star, if it is confirmed to be a planetary-mass companion, we estimate a mass of 3--5\,\mj and a separation of 730 $\pm$ 10\,au for it, a low-mass planet similar to 51~Eri~b and AF~Lep~b and also one of the widest-orbit planets imaged so far.

\begin{acknowledgements}
We thank Rico Landman, Ben Jia, Richelle van Capelleveen, Zephyr Penoyre and Anthony G.A. Brown for their valuable discussion. 
Based on observations collected at the European Organisation for Astronomical Research in the Southern Hemisphere under ESO programmes 0101.C-0153(A), 108.220Q.001 and 110.23NY.001.
Part of this research was carried out in part at the Jet Propulsion Laboratory, California Institute of Technology, under a contract with the National Aeronautics and Space Administration (80NM0018D0004).
This research has used the SIMBAD database, operated at CDS, Strasbourg, France \citep{Wenger2000}. 
This research made use of \textsc{Astropy}, a community-developed core \textsc{Python} package for Astronomy \citep{Astropy2013,Astropy2018}, \textsc{SciPy} \citep{Scipy2020}, \textsc{NumPy} \citep{Numpy2020} and \textsc{Matplotlib}, a Python library for publication quality graphics \citep{Matplotlib2007}.
This research made use of \textsc{Lightkurve}, a Python package for Kepler and TESS data analysis \citep{lightkurve2018}.
This work has benefitted from The UltracoolSheet, maintained by Will Best, Trent Dupuy, Michael Liu, Rob Siverd, and Zhoujian Zhang, and developed from compilations by \cite{Dupuy2012, Dupuy2013, LiuM2016, Best2018, Best2020}.
This research made use of the Montreal Open Clusters and Associations (MOCA) database, operated at the Montr\'eal Plan\'etarium (J. Gagn\'e et al., in prep)
We thank the writers of these software packages for making their work available to the astronomical community.
\end{acknowledgements}

\bibliographystyle{aa}
\bibliography{main}

\begin{appendix} 
\onecolumn
\section{Central star images}
\label{app:starims}
\begin{figure}[h!]
\centering
  \includegraphics[width=0.8\textwidth]{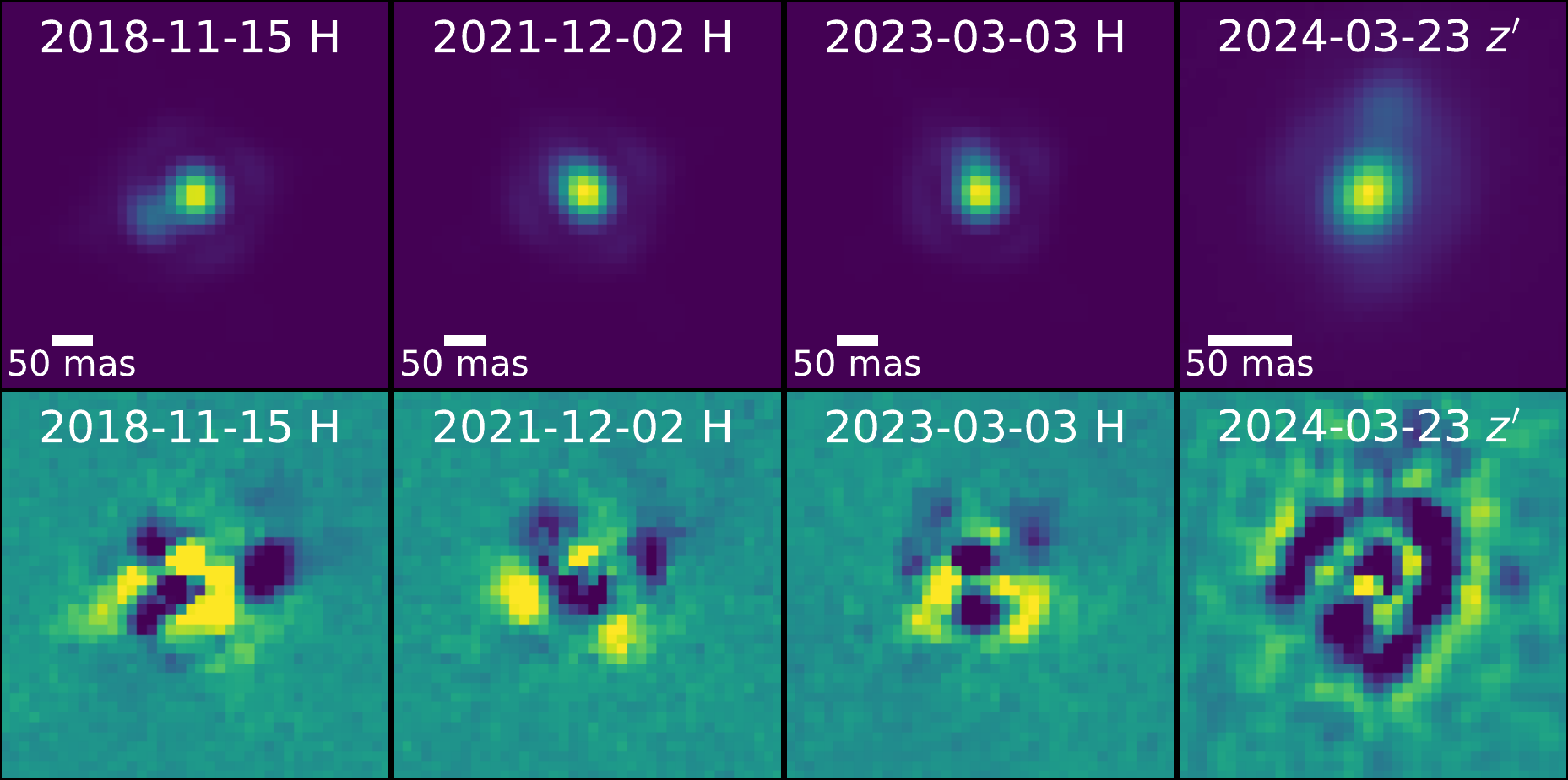} 
\caption{Images of the central stars from 2018-11-15 to 2024-03-23. All images are rotated to the direction where North is up and East is left. Upper row: star images before subtraction; Lower row: residuals after fitted model subtraction.}
\label{fig:starims}
\end{figure}

The images of the central stars from 2018-11-15 to 2024-03-23 in $H$ and $z^\prime$ bands are presented in Fig.~\ref{fig:starims}, including the model subtracted residuals.

\onecolumn
\section{Posterior distribution of SED fitting for the central two stars}
\label{app:sed_post}
\begin{figure}[h!]
\centering
  \includegraphics[width=0.9\textwidth]{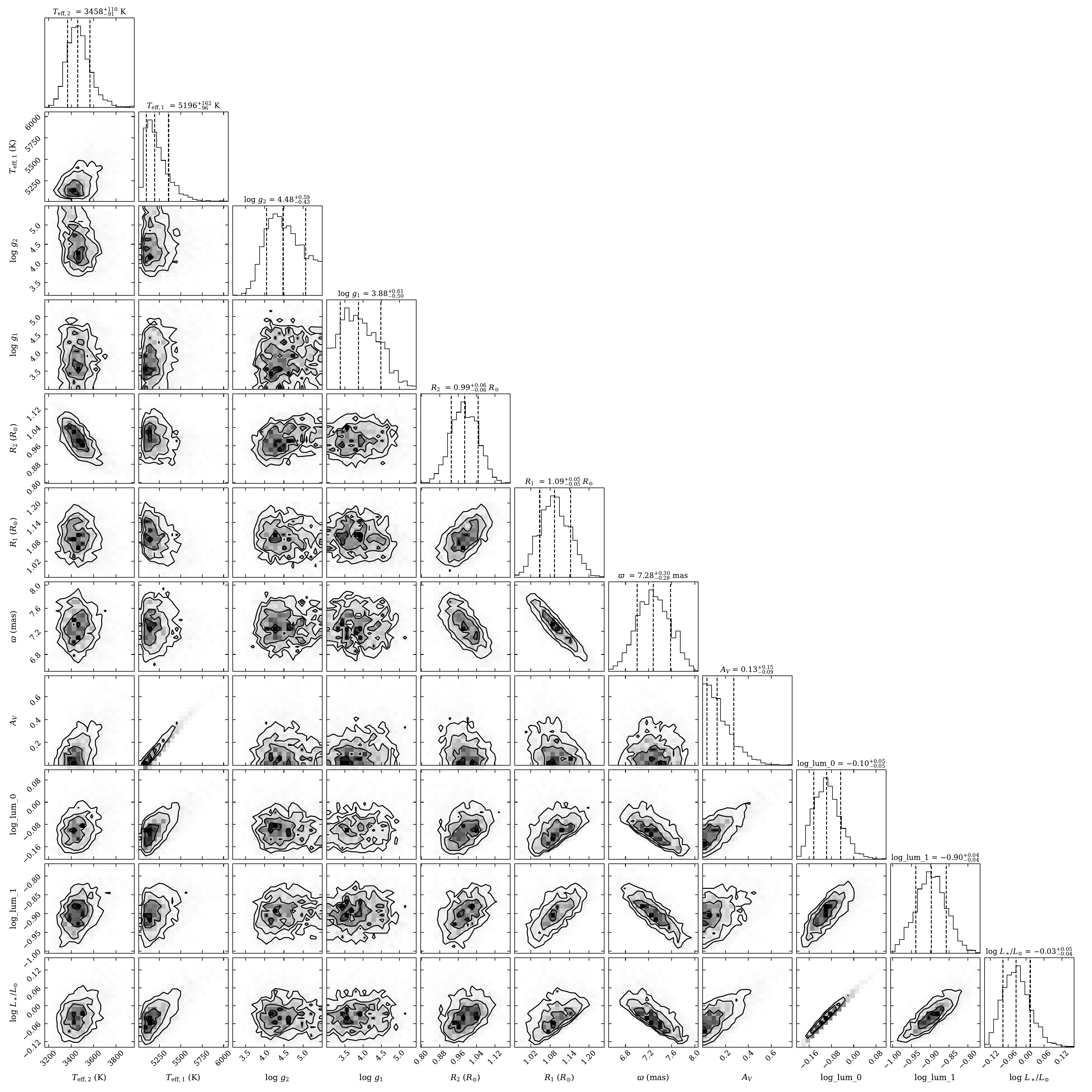} 
\caption{Posterior distributions of the SED fitting parameters for the central stars.}
\label{fig:sed_con_star_bound}
\end{figure}

The posterior distributions of SED parameter fitting of the central two stars are presented in Fig.~\ref{fig:sed_con_star_bound}.

\onecolumn
\section{Radial velocity injection test}
\label{app:rvinj}
\begin{figure}[h!]
\centering
  \includegraphics[width=0.7\textwidth]{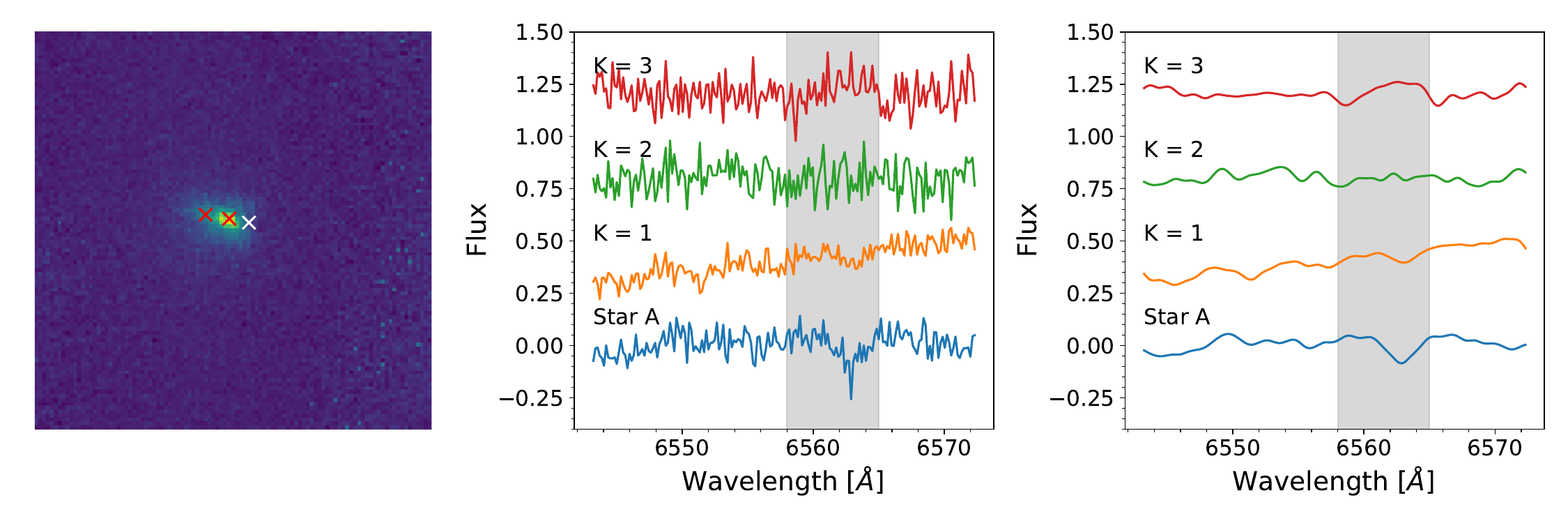} 
  \includegraphics[width=0.7\textwidth]{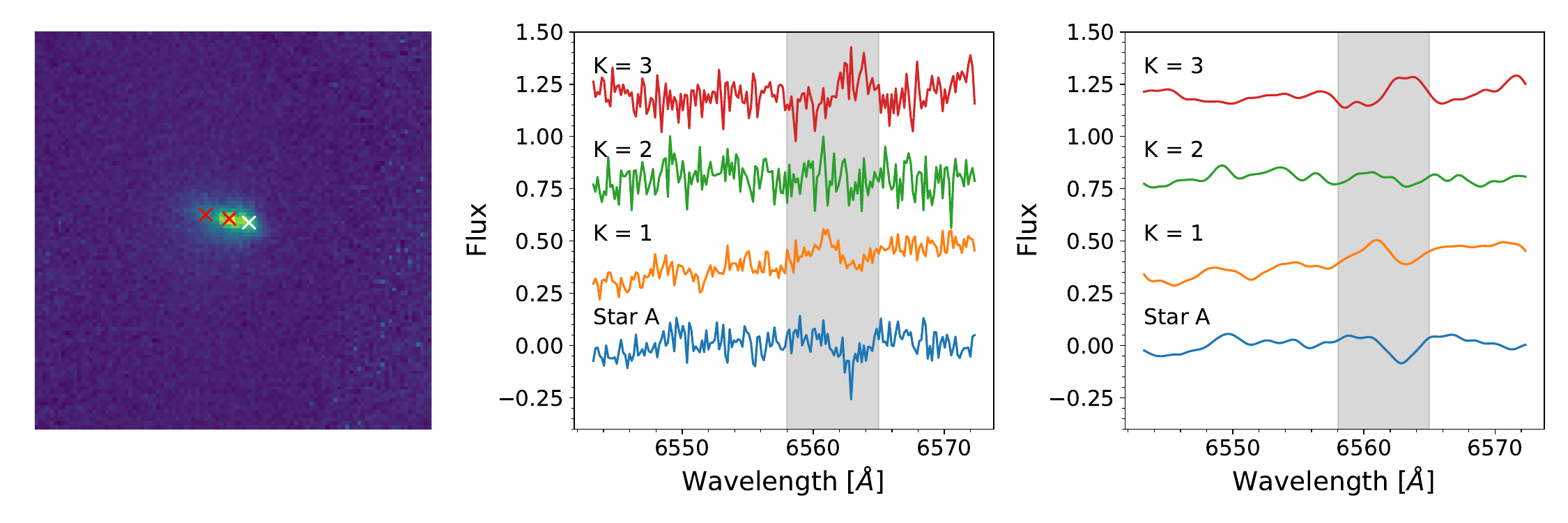} 
\caption{Injection test for radial velocity measurements in VIS-X IFU data. Upper panel: original data; lower panel: a fake Gaussian signal of the same contrast and same blue shift as the fainter star is injected at the same separation but in the opposite direction shown by the cross marker in white. Before the injection, there is no signal in the first component (K=1) in the upper panel. After the injection, we detected the emission at the supposed blue shift and amplitude in the first component (K=1) in the lower panel.}
\label{fig:rv_injection}
\end{figure}

We injected artificial signals in the VIS-X IFU data to verify the PCA method we developed in Sect.\ref{sec:radial_velocity}. 
We fit a 2D Gaussian model to the primary star and took the fitted model as the PSF template. Then we created a 1D Gaussian signal with an amplitude of 0.2 as the injected spectrum which is of similar brightness to the fainter star. We scaled the PSF template with the 1D Gaussian signal and injected them at the same separation and blue shift as the fainter star but in the opposite direction. We retrieved the signal successfully using the same PCA method we developed for the fainter star as demonstrated in Fig.\ref{fig:rv_injection}.

\section{Relative astrometry of all sources}
\label{app:relative_astroall}
\begin{figure}[h!]
\centering
  \includegraphics[width=0.38\textwidth]{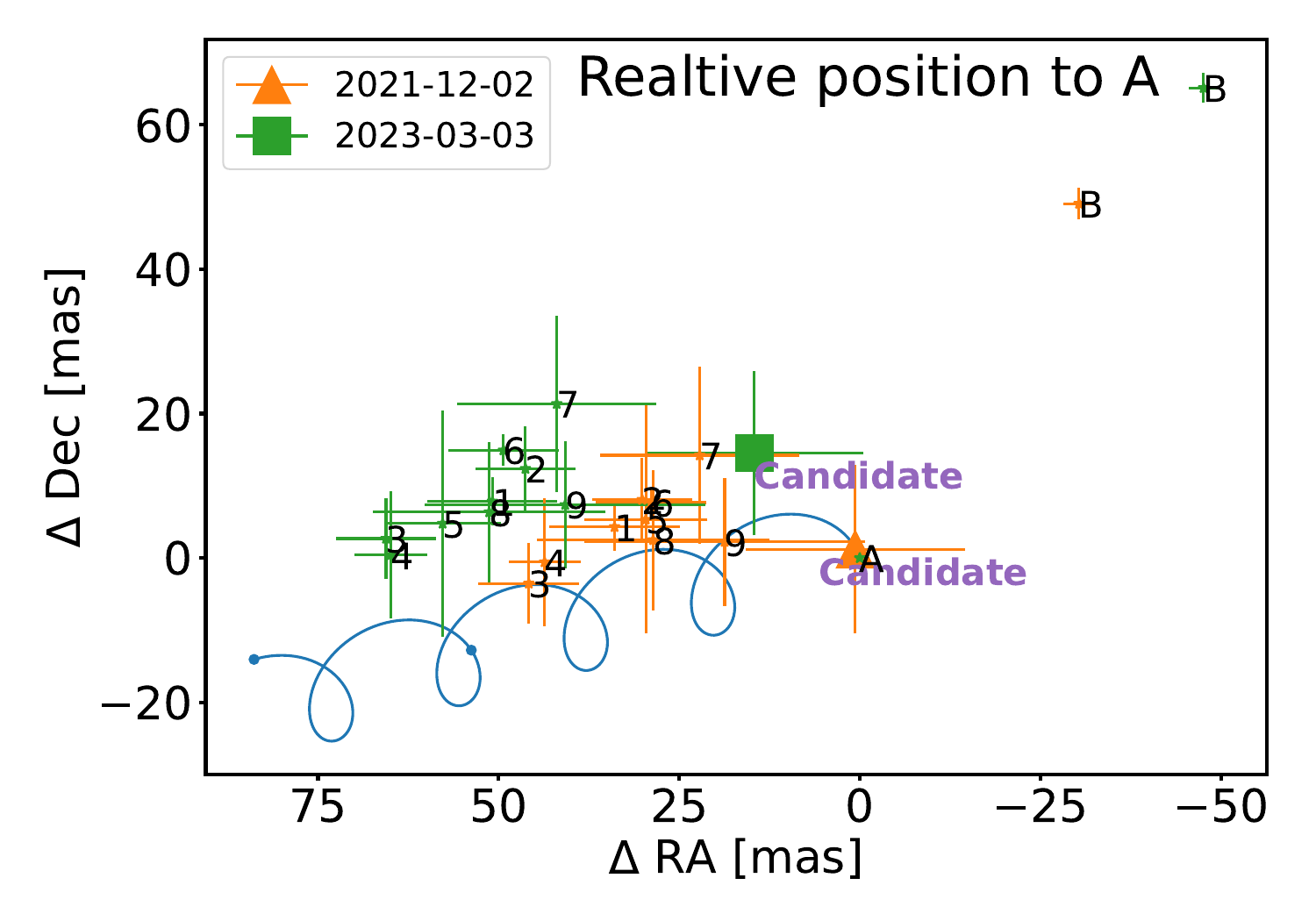} 
  \includegraphics[width=0.4\textwidth]{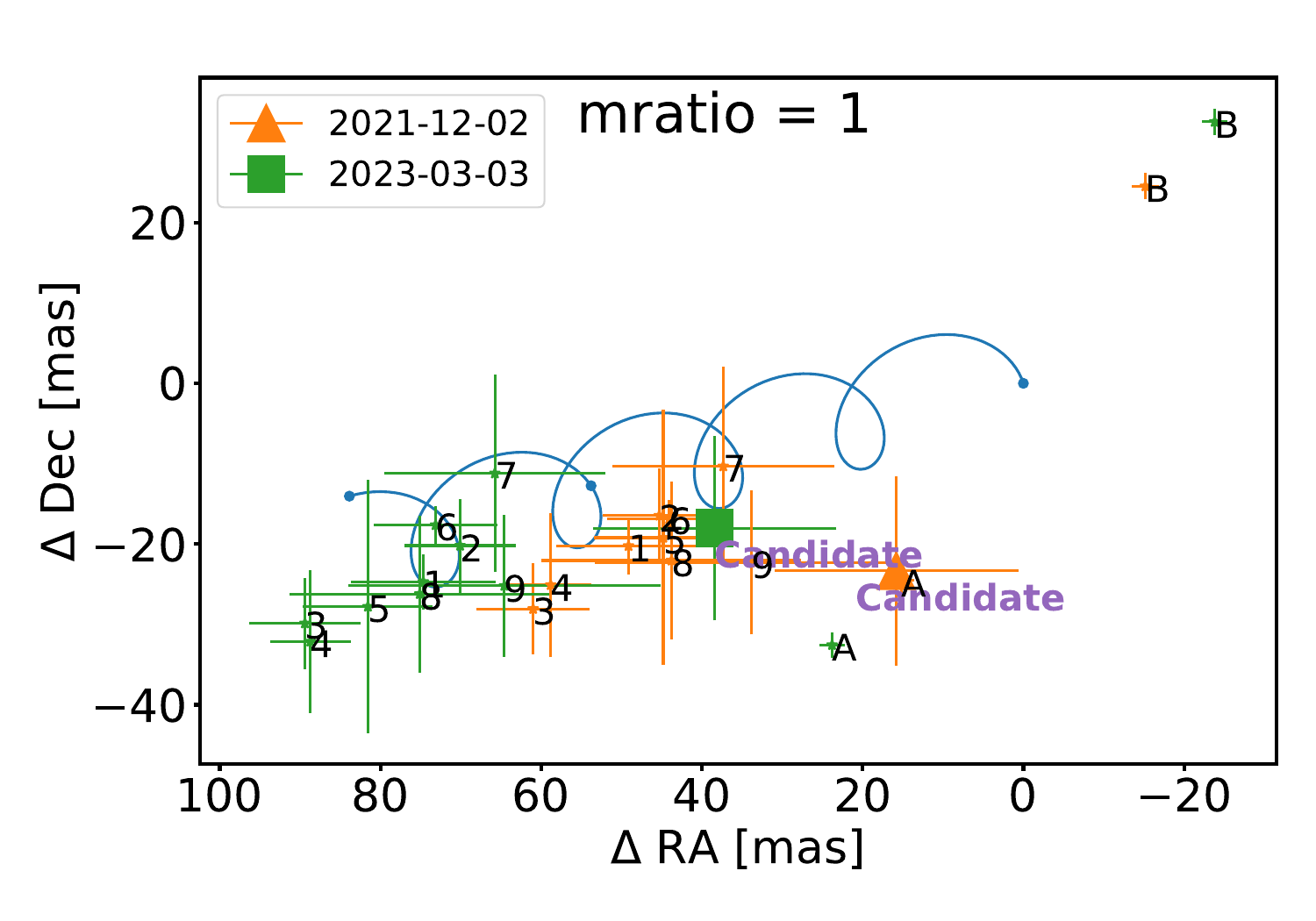} 
\caption{Relative astrometry all sources in RA and Dec offsets. 
Left panel: relative astrometry of all sources to the primary star (mass ratio = 0).
Right panel: relative astrometry of all the sources to the barycentre when the mass ratio = 1.
The positions of all sources of the first epoch are calibrated to the origin, so the positions of the later epochs are subtracted by the positions of the first epoch.
The candidate is shown by the triangle and square markers with the epoch 2021-12-02 in orange and the epoch 2023-03-03 in green.
The sources are labelled with the same number in Fig.~\ref{fig:wholeime} using the same colour convention for epochs.
The blue line is the trajectory of a static background star evolving from 2018-11-15 and the two points on the track are the positions on 2021-12-02 and 2023-03-03 calculated by the proper motion and parallax reported in Gaia.}
\label{fig:astrometry_all}
\end{figure}

Figure.~\ref{fig:astrometry_all} shows the relative astrometry of all sources in FoV when the mass ratio = 0 and 1. Regardless of the mass ratio of the central stars, the candidate has the most significant motion compared to other sources except for the central stars, and it deviates from the proper motion cluster of other sources.

\end{appendix}

\end{document}